\documentclass[8pt]{article}
\usepackage[utf8]{inputenc}
\usepackage[english]{babel}
\usepackage{amsmath}
\usepackage{amsfonts}
\usepackage{amssymb}
\usepackage{amsthm}
\usepackage{graphicx}

\usepackage[a4paper, left=2.5cm, right=2.5cm, top=2cm, bottom=2cm]{geometry}
\usepackage{xcolor}
\usepackage{hyperref}
\usepackage{doi}
\usepackage{cite}

\newtheorem{theorem}{Theorem}
\newtheorem{lemma}{Lemma}
\newtheorem{statement}{Statement}
\newtheorem{remark}{Remark}
\newtheorem{corollary}{Corollary}

\title{Cauchy problem for the localized wave propagation in continuous model of the one-dimensional diatomic crystal\footnote{This work was done under the financial support of FAPERJ APQ1 number E-26/210.614/2024.}}
\author{Sergey Sergeev\footnote{Federal University of Rio de Janeiro, Institute of Mathematics, Department of Applied Mathematics, Rio de Janeiro, Brazil, sergeevse1@im.ufrj.br}}
\date{}

\begin{document}
\maketitle

\begin{abstract}
We study the continuous approximation model of the localized wave propagation corresponding to the one-dimensional Cauchy problem for the diatomic crystal lattice. This approximation is written as a system of two pseudo-differential equations with localized initial conditions. We assume two small parameters in this formulation --- the lattice step and the size if the initial perturbation. We construct the asymptotic solution of the continuous Cauchy problem with respect to the size of perturbation. 

The ratio of the small parameters significantly affects the form of the solution. We consider two situations --- when the size of the perturbation is sufficiently large and when it is quite narrow and comparable with the lattice step. In each situations we provide analytical formulae for the asymptotic solution via Airy function. 



\end{abstract}

\section{Introduction}
The polyatomic lattice models are very important from the point of view of applications and fundamental studies and the investigation  of such models is continuing up today  \cite{Xiong, Ma, Kesav, Ferreyra}. The one-dimensional diatomic crystal model is a fundamental model in solid state physics \cite{brillouin_wave_1953, born_dynamical_2002, kittel_introduction_nodate, Vonsovskij, askar_lattice_1985} and it appears in very different applications, for example in the studies of wave propagation in composite materials \cite{Carta}, wave propagation in electric lines \cite{brillouin_wave_1953, vega_computer-aided_1997, Lin}. In the bio-physics and DNA modeling the discrete model leads to the linear acoustical equation for the in-phase oscillations  and non-linear model for the optical out-phase oscillations \cite{Peyrard,  Okaly}. The weakly non-linear  diatomic lattice was also considered in \cite{porubov_nonlinear_2013}.

We are considering the linear diatomic model. It consists of two types of atoms with different masses $m_1$ and $m_2$ (where $m_1 > m_2$). The atoms arranged in an alternating pattern such that each atom is surrounded by atoms of the opposite type. In equilibrium, the atoms located in the lattice points with the nearest-neighbor distance $d = a/2$, where $a$ is the length of the atomic cell. 

In the presence of some perturbation, the atoms oscillate about their equilibrium positions. We denote the displacement of heavy atoms (mass $m_1$) as $u_{2k}(t)$ and light atoms (mass $m_2$) as $v_{2k+1}(t)$. 
The dynamic under nearest-neighbor interactions is governed by the coupled equations following from the Newton's second law.

Let us present the equations of the oscillations in the dimensionless form. To do so, we introduce the characteristic distance of the wave propagation $L$, then we can introduce the  dimensionless lattice step
$h=d/L\ll 1$. The smallness of $h$ means that the distance of the wave propagation is much bigger than the distance between the nearest points. With the help of the parameter $h$, we  define the lattice $x_n=nh$, $n\in\mathbb{Z}$. Let us introduce in  addition the characteristic velocity $c$ of the waves and the characteristic time scale $T=L/c$. 

The dimensionless system of equations on the lattice $x_n=nh$ has the form
\begin{gather}
\label{osc_lattice_dimens}
h^2 \ddot{u}_{2k}(t)=\gamma_1(v_{2k-1}(t)-2u_{2k}(t)+v_{2k+1}(t)),\\
\nonumber
h^2 \ddot{v}_{2k+1}(t)=\gamma_2(u_{2k}(t)-2v_{2k+1}(t)+u_{2k+2}(t)).
\end{gather}
Here $\gamma_{1,\,2}=K d^2/(m_{1,\,2} c^2)$, $\gamma_1<\gamma_2$, and a constant $K$ represents the spring constant characterizing the interatomic force. 

A key characteristic of the diatomic model is the existence of two distinct frequency branches for propagating waves. These branches are separated by a frequency gap and define the acoustic and optical wave modes.

Despite the fundamental importance of the diatomic model, surprisingly little attention has been paid to the associated Cauchy problem for either the discrete or continuous formulations. In the present work we focus ourselves on the asymptotic analysis of the Cauchy problem for continuous analogue to the discrete system \eqref{osc_lattice_dimens}. In this formulation two questions arise - construction of the continuous model and construction of the corresponding initial data for this model.  

The transition from discrete lattice equations to continuous models has been studied in various contexts. Typically, such analyses consider the limit passing as the lattice step approaches zero. In the studies of the equations on the lattice, and in particularly in the diatomic lattice model, the approach is based on the introduction of the shift operators. This can be done in the form of the pseudo-differential operators \cite{Andrianov, Fusi} or by the classical substitution of the plain wave \cite{brillouin_wave_1953, born_dynamical_2002, kittel_introduction_nodate, Vonsovskij, askar_lattice_1985}. Then the assumption of the smallness lattice step is used and thus the approximate system of equations derived. One may call it, following \cite{Gomez}, the standard procedure. In the principal term we obtain the wave equation for the acoustical wave mode.

The problems arise when one needs to take into account the correction to the principal terms. For example, the correction to the  wave equation leads to the appearance of the 4-th derivatives and the dispersion effects. At the same time the new approximate equation becomes ill-posed, since the symbol of the operator can be negative. In order to avoid this problem, another differential equation is constructed with the help of the Pade approximation and the resulting equation contains mixed derivatives in time and space. One may call it nonstandard procedure \cite{Gomez, Andrianov}. Notice, that the original system itself is well-posed and the mentioned above problem is just the incorrect approximation. 

We propose the general approach, based on the pseudo-differential operators \cite{maslov_fedoruk_1976, MasDan86_1, Botchway}. In this way we do not start from the approximation of the symbol of the equation, but use it as a whole. We write the shift operator on the lattice as a pseudo-differential, which leads to the system of two pseudo-differential equations. This approach preserves the complete dispersion relations for both branches, not only their long-wave approximations, providing the well-posed system of the equations.

The next step in our study is the posing of the initial perturbation and consideration of the Cauchy problem. The Cauchy problem itself rises various problems, starting from the continuous analogues of the initial perturbation on the lattice, especially for the diatomic system, when one has to separate the perturbations of atoms of different kind.

As previously noted, the Cauchy problem has received comparatively little attention, most of the literature focusing solely on the equations themselves. In \cite{blanc_newton_2012} the limit equations, together with the corresponding initial data, were derived for uniform one-dimensional lattices in both the linear and nonlinear cases. Discrete and continuous models of uniform lattice oscillations for localized initial data were studied in \cite{Baimova}. In that work the evolution and energy dispersion of the solution were examined for both discrete and continuous models; the initial conditions were taken to be a wave packet with compactly supported amplitude, referred to as a localized wave packet. In \cite{musgrave_shock_1976} the propagation of shock waves in semi-infinite diatomic chains was investigated under trivial initial conditions, but continuous equations were not obtained. Here we propose a natural approach, from the standpoint of pseudo-differential operators, for treating the initial conditions by means of the Kotel’nikov-Whittaker-Shannon interpolation.

Following the ideas of \cite{dobrokhotov_propagation_2017, sergeev_asymptotic_2019, grushin_asymptotics_2020}, we introduce the notion of localized initial conditions, which differs from that of \cite{Baimova}. We do not assume that the initial data have compact support; rather, we require only that they decay rapidly at infinity. This leads to the introduction of a new parameter - the width of the initial perturbation - which governs the properties of the propagating wave and determines the appropriate form of the approximate equation. In contrast to \cite{Baimova}, where the support of the amplitude is assumed to be sufficiently large (depending on the frequency of the wave packet), our approach treats the localization width as an independent parameter. This makes the framework considerably more flexible and allows a wider range of asymptotic regimes to be considered.

Depending on the ratio between the localization parameter and the lattice spacing, the nonlinear dispersion relation should be considered. When these two quantities are related by a power law of order $3/2$, dispersion effects become significant. In this connection we mention the work of \cite{berezhnyy_continuum_2006}, where the continuum limit for a large number of particles in three dimensions was examined and a mesoscopic parameter obeying the same $3/2$ power law was introduced, which seems the natural relation for the appearance of the dispersion.

We consider two distinct types of perturbation. In the first, the lattice perturbation is described by a smooth function; we call this an in-phase initial perturbation. In the second, the perturbation is described by a rapidly oscillating function and is termed an out-of-phase initial perturbation.

We also examine the regime in which the size of the initial perturbation is comparable to the lattice spacing; we refer to this as a narrow perturbation. In this regime the propagating waves exhibit qualitatively different behaviour dominated by strong dispersion. We derive asymptotic formulae that describe the propagation of the acoustic wave under these conditions.

An asymptotic analysis of the Cauchy problem is carried out by means of semiclassical analysis and Maslov’s canonical operator \cite{maslov_fedoruk_1976}. The analysis yields explicit analytical asymptotic formulae, expressed in terms of Airy functions and their derivatives, that describe the wave propagation. These formulae are computationally inexpensive compared with the direct numerical solution of a large system of ODEs. 

The presence of the analytical formulae allows us to provide the direct analysis of the waves, such as definition of a Gouy time-shift phase for the optical mode, analogous to the classical Gouy phase shift that appears in the spatial variables \cite{Boyd, Feng}.  In addition, we obtain a well-posed differential equation that is considerably simpler than the corresponding Pade approximation.

This work is organized as follows.  Section \ref{sec_main_eq_init_cond} formulates the continuous model, initial conditions, and poses the Cauchy problem. We also discuss in this section the validity of the application of the Kotel'nikov-Whittaker-Shannon interpolation formulas.  Section \ref{sec_small_delta} is dedicated to the sufficiently large perturbation regime, while in Section \ref{sec_strong_disp} we treat the narrow perturbations (with detailed proofs in Appendix \ref{app_stron_disp}). Section \ref{sec_exmpl} provides example for a perfect diatomic lattice illustrating both regimes.

Author thanks Pavel S. Petrov and Anatoly Yu. Anikin  for the useful discussions and valuable comments.

\section{Definition of the initial conditions and main equations.}
\label{sec_main_eq_init_cond}

\subsection{Continuous equations}

Our first goal is to rewrite  equations (\ref{osc_lattice_dimens}) in the continuous form. We follow the standard procedure rewriting the lattice equations in with the help of the shift operators
$$
T_hf(x)=f(x+h)=e^{h\frac{\partial}{\partial x}}f(x),\quad T^{-1}_hf(x)=f(x-h)=e^{-h\frac{\partial}{\partial x}}f(x).
$$
The last equalities can be verified, for example, by the Taylor expansion with respect to the small parameter $h$.

We introduce two continuous functions $u(x,\,t)$ and $v(x,\,t)$ and write the equations (\ref{osc_lattice_dimens}) for these functions in the points $x_{2k}$ and $x_{2k+1}$. 
We assume that the equations hold for any $h\in(0,\,h_0]$ for some $h_0>0$ fixed. After that we can write equations (\ref{osc_lattice_dimens}) in the point $x\in\mathbb{R}$
\begin{align}
\label{u_cont_eq}
\begin{split}
h^2 \ddot{u}(x,\,t)=\gamma_1(T^{-1}_hv(x,\,t)-2u(x,\,t)+T_hv(x,\,t)),\\
h^2 T_h\ddot{v}(x,\,t)=\gamma_2(u(x,\,t)-2T_hv(x,\,t)+T^2_hu(x,\,t)).
\end{split}
\end{align}
These equations coincide with the equations (\ref{osc_lattice_dimens}) when $x=x_{n}$. Let  $u(x,\, t)$ and $v(x,\, t)$ be the solution of the these equations, then the restriction $u(x_{n},\,t)$, $v(x_{n},\,t)$ onto the lattice gives us the description of the solution of the system (\ref{osc_lattice_dimens}).

 We can rewrite the equations (\ref{u_cont_eq}) in the form of the system of two pseudo-differential equations with the help of the following equality
$$
T_h+T_h^{-1}=2\cos(hD),\quad D=-i\frac{d}{dx}.
$$
Let us denote $U(x,\,t)=\bigl(u(x,\,t),\, v(x,\,t)\bigr)^T$ the vector-function of the two unknown functions, then the equations (\ref{u_cont_eq})  take the form
\begin{equation}
\label{main_sys_cont}
-h^2\frac{\partial^2}{\partial t^2}U(x,\,t)=2 \Gamma L(hD)U(x,\,t),
\end{equation}
where 
\begin{align}
\label{main_operator}
L(hD) &= \left(
\begin{array}{cc}
1 & -\cos(hD)\\
-\cos(hD) & 1
\end{array}
\right),\quad
\Gamma=\left(
\begin{array}{cc}
\gamma_1 & 0\\
0 & \gamma_2
\end{array}
\right),\quad 0<\gamma_1<\gamma_2.
\end{align}
We treat the system (\ref{main_sys_cont}), (\ref{main_operator}) as a continuous analogue of the equations on the lattice (\ref{osc_lattice_dimens}). These equations are our main equations which we will study. 

The operator \eqref{main_operator} is the matrix pseudo-differential operator. Let us describe how this operator acts on the vector function $U(x,\,t)=(u(x,\,t),\,v(x,\,t))$. Following \cite{Botchway, MasDan86_1 } we define the action of this operator via semi-discrete Fourier transform \cite{trefethen_spectral_2000} and the Kotel'nikov-Whittaker-Shannon interpolation formulas  \cite{MasDan86_1, higgins_sampling_1996, Andrianov}.

Let us define the vector $W(x)=(u(x), v(x))^T$ of two functions $u(x)$ and $v(x)$ and then consider its restriction $W_n=(u(nh), v(nh))$ on the lattice $x_n=nh$, $n\in\mathbb{Z}$. Let us denote $\widetilde{W}(p)$ its semi-discrete Fourier transform  
$$
\widetilde{W}(p)=\sum\limits_{n\in\mathbb{Z}}W_n e^{-i p n},\quad p\in [-\frac{\pi}{2h},\,\frac{\pi}{2h}].
$$

Let $M(T^{\pm 1}_h)$ be the $2\times 2$ matrix with its elements depend on the shift operators $T_h^{\pm 1}$. The action of this matrix is defined as follows
\begin{equation}
\label{p_semi_Fourier}
M(T^{\pm 1}_h)  W(x)=\frac{h}{\pi }\int\limits_{-\pi/(2h)}^{\pi/(2h)}M(e^{\pm i p})\widetilde{W}(p)e^{\frac{i}{h}xp}dp.
\end{equation}
This equality can be verified by direct computation because of the linearity and the equality  $T_h^{\pm 1}e^{\frac{i}{h}px}=e^{\pm ip}e^{\frac{i}{h}px}$, meaning that $e^{\pm ip}$ is the symbol of the shift operators.

The formula (\ref{p_semi_Fourier}) allows us to define the action of the pseudo-differential operator $L(hD)$ on the vector-function $U(x,\,t)$ via the product $L(p)\widetilde{U}(p,\,t)$. The matrix $L(p)$ is called the symbol of the operator $L(hD)$.

\subsection{Definition of the initial conditions.}

At the initial time moment, the atoms are subject to the localized perturbation with non-zero velocity. We suppose that both initial conditions can be described by the values $W_{n}$, $n\in\mathbb{Z}$, obtained from the following function
\begin{equation}
\label{Gauss_pack_lattice}
W(x)=A(\frac{x}{\mu})e^{\frac{i}{\mu} k_0 x},\quad W_n=W(n h),
\end{equation}
where $\mu$ is the length of the localization of the perturbation and $k_0$ is the frequency of the wave packet. Amplitude $A(\xi)$ is assumed to be in the Schwartz space. In particular we will consider the case of the Gaussian exponential $A(\xi)=e^{-\xi^2/2}$.

The localization parameter $\mu$ is defined as follows
$$
\mu=\frac{\ell}{L}\ll 1,
$$
where $\ell$ is the characteristic diameter of the perturbation in the dimensional lattice and the $L$ is the distance of the wave propagation. 

The smallness of $\mu$ shows that the initial perturbation is localized with respect to the propagation distance of the waves. This parameter plays an important role in our asymptotic analysis, since such analysis will be done with the respect of the smallness of this parameter.

We have two different small parameters: $h$ which corresponds to the lattice step and $\mu$ which corresponds to the length of the initial perturbation. Let us introduce the following ratio
$$
\delta=\frac{h}{\mu}\equiv \frac{d}{\ell}.
$$
Despite the parameters $\mu$ and $h$ are small, the ratio $\delta$ can be quite small or not. This depends on the size of the initial perturbation. If $h\ll \mu$, then also  $\delta\ll 1$ and we say that this case corresponds to the sufficiently long waves with respect to the lattice step.

On the other hand, if  $\mu$ is comparable with $h$, then $\delta<1$ but may not very small. We say that this is the case of the short waves with respect to the lattice step or the narrow perturbation.

Our goal is to separate the perturbations of the atoms of the first kind and of the second kind and present the continuous form for such perturbations. Here we provide the general approach which is valid for both initial perturbation and initial velocity.

Recall that if the number of the lattice point $n$ is even, then the $n$-th atom on the lattice is of the first kind and the odd numbers correspond to the atoms of the second kind. Let the initial data for the atoms is of the form
$$
u_{2k}(0)=W_{2k}=W(2k\delta),\quad
v_{2k+1}(0)=W_{2k+1}=W(2k\delta+\delta).
$$

Since the action of the operator is defined via semi-discrete Fourier transform and the Kotel'nikov-Whittaker-Shannon interpolation, we use the same approach in order to define the continuous form of the initial perturbation. We have

\begin{equation}
\label{tilde_W1_W2}
\widetilde{W}_1(p)=\sum\limits_{k\in\mathbb{Z}}W(2k\delta)e^{-\frac{i}{\mu}p 2kh},\quad
\widetilde{W}_2(p)=\sum\limits_{k\in\mathbb{Z}}W(2k\delta+\delta)e^{-\frac{i}{\mu}p (2k+1)h},
\end{equation}
where $p$ is in the interval
$$
p\in B_1\equiv\left[-\frac{\pi}{2\delta},\,\frac{\pi}{2\delta}\right].
$$
This interval $B_1$ is called the Brillouin zone. Let us denote as $l=\pi/\delta$ the length of  $B_1$, then we have 
$\widetilde{W}_1(p+l)=\widetilde{W}_1(p)$ and $\widetilde{W}_2(p+l)=-\widetilde{W}_2(p)$ for $p\in B_1$. 

The Kotel'nikov-Whittaker-Shannon interpolation gives the following localized functions
$$
W_1\left(\frac{x}{\mu}\right)=\frac{\delta}{\pi }\int\limits_{B_1}\widetilde{W}_1(p)e^{\frac{i}{\mu}xp}dp,\quad W_2\left(\frac{x}{\mu}\right)=\frac{\delta}{\pi }\int\limits_{B_1}\widetilde{W}_2(p)e^{\frac{i}{\mu}xp}dp.
$$

We apply this approach for the initial data, which gives us the initial data for each type of atoms
\begin{equation}
\label{init_vec}
U^0\left(\frac{x}{\mu}\right)=\Biggl(W_1\left(\frac{x}{\mu}\right),\,W_2\left(\frac{x}{\mu}\right)\Biggr)^T,\qquad
U^1\left(\frac{x}{\mu}\right)=\Biggl(W^1_1\left(\frac{x}{\mu}\right),\,W^1_2\left(\frac{x}{\mu}\right)\Biggr)^T,
\end{equation}
where the vector $U^{0}(x/\mu)$ corresponds to the initial perturbation of the position and $U^{1}(x/\mu)$ is the initial velocity.

Another parameter of the initial perturbation \eqref{Gauss_pack_lattice} is the frequency $k_0$. 
Depending on the value of the frequency $k_0$ we obtain either smooth in-phase localized signal for $k_0\sim 0$ or oscillating signal, which modeling the out-of-phase localized signal for $k_0\sim \pi/\delta$.

When $k_0=\pi/\delta$, the restriction of the exponential $e^{i k_0 x/\mu}$ on the lattice $x=nh$ gives $e^{i k_0 n h/\mu}=(-1)^n$. It means that atoms of the different kind have the opposite perturbation which gives the smooth out-of phase initial perturbation
$$
W_{2k}=A(2k\delta) e^{i \xi_0 2k\delta},\quad W_{2k+1}=-A((2k+1)\delta)e^{i \xi_0 (2k+1)\delta}.
$$
This gives us two smooth continuous functions such that that $W_1(x/\mu)=-W_{2}(x/\mu)$.

\subsection{Application of the Shannon interpolation to the localized functions.}
The Kotel'nikov-Whittaker-Shannon interpolation formula gives an exact approximation for the band-limited functions functions $f(x)$, meaning that the support of the Fourier transform is inside the Brillouin zone. In our case of the localized initial perturbation \eqref{Gauss_pack_lattice} is not band-limited. On the other hand the Fourier transform of the function from the Schwartz space is also belongs to the Schwartz space. For such localized initial perturbation we can give lower bound for the localization, when the interpolation formula gives only small correction. 

We introduce the $\mu$-Fourier transform, which is
$$
\widehat{W}(p)=F_\mu[W(x/\mu)]=\frac{1}{\sqrt{2\pi} \mu}\int\limits_{\mathbb{R}}W\left(\frac{x}{\mu}\right)e^{-\frac{i}{\mu}px}dx.
$$
The inverse $\mu$-Fourier transform gives the localized function
$$
F^{-1}_\mu[\widehat{W}(p)]=\frac{1}{\sqrt{2\pi}}\int\limits_{\mathbb{R}}\widehat{W}(p)e^{\frac{i}{\mu}px}dp=W\left(\frac{x}{\mu}\right).
$$

In \cite{Brown, Butzer} the estimation for the non-band-limited functions was proved. Following these works, we have  the following estimation.
\begin{statement}
\label{st_Kot_correc}
Let function $W(x)$ be the wave-packet \eqref{Gauss_pack_lattice} and let $k_0\in[0,\,\pi/2\delta]$, then we have the following estimation for the Kotel'nikov-Whittaker-Shannon interpolation
\begin{equation}
\label{aliasing_error}
\left|W(\frac{x}{\mu})-\frac{\delta}{\pi}\int\limits_{-\pi/2\delta}^{\pi/2\delta}\widetilde{W}(p)e^{\frac{i}{\mu}px}dp
\right|\le \sqrt{\frac{2}{\pi}}\int\limits_{|p|\ge \pi/2\delta}|\widehat{W}(p)|dp.
\end{equation}
\end{statement}
Estimation \eqref{aliasing_error} shows, that if the spectrum of the function $W(\xi)$ is localized, then we still can use the Shannon interpolation, providing the error is small. This allows us to determine the localization parameter $\mu$ when the interpolation gives small correction, for fixed lattice step $h$ and frequency $k_0$. Let us choose, for example, in \eqref{Gauss_pack_lattice} the initial amplitude in the form of the Gaussian exponential $A(\xi)=e^{-\xi^2/2}$, then we arrive to the following estimation
\begin{equation}
\label{aliasing_error_Gauss}
\left|W(\frac{x}{\mu})-\frac{\delta}{\pi}\int\limits_{-\pi/2\delta}^{\pi/2\delta}\widetilde{W}(p)e^{\frac{i}{\mu}px}dp
\right|\le erfc\left(\frac{1}{\sqrt{2}}\left(\frac{\pi}{2\delta}-k_0\right)\right)+erfc\left(\frac{1}{\sqrt{2}}\left(\frac{\pi}{2\delta}+k_0\right)\right).
\end{equation}
The classical order of the error is $O(\mu^2)$ in the semiclassical analysis and for $k_0=0$ we have the following equation to determine the lower bound for the parameter $\mu$:
\begin{equation}
\label{low_bound_mu_eq}
2erfc\left(\frac{\pi}{2\sqrt{2}}\frac{\mu}{h}\right)=\mu^2.
\end{equation}

In the following Lemma we establish more profound connection between the $\mu$-Fourier transform $\widehat{W}(p)$ of the function $W(x/\mu)$ and the semi-discrete Fourier transforms $\widetilde{W}_{1,\,2}(p)$.
\begin{lemma}
\label{lm1_Poisson}
Let function $W(x)$ is defined in \eqref{Gauss_pack_lattice} and let $k_0\in[0,\,\pi/\delta]$ and  $\delta=h/\mu\ll1$. Let $\widehat{A}(p)$ be the Fourier transform of the function $A(x)$. Then, uniformly for $p\in B_1$,  the following relations hold 
\begin{gather}
\label{tilde_W_1_Poisson}
\widetilde{W}_1(p)=\frac{1}{\delta}\sqrt{\frac{\pi}{2}}\left(\widehat{A}(p-k_0)+\widehat{A}(\frac{\pi}{\delta}+p-k_0)\right)+O(\delta^\infty),\\
\label{tilde_W_2_Poisson}
\widetilde{W}_2(p)=\frac{1}{\delta}\sqrt{\frac{\pi}{2}}\left(\widehat{A}(p-k_0)-\widehat{A}(\frac{\pi}{\delta}+p-k_0)\right)+O(\delta^\infty).
\end{gather}
\end{lemma}

\begin{proof}
The proof is based on the Poisson summation formula \cite{grushin_asymptotics_2020, stein_introduction_1975, henrici_applied_1986_vol2}, which says that for any $f(x)\in L_1(\mathbb{R})$ we have the following equality
\begin{equation}
\label{Poisson_x_zero}
\sum\limits_{k\in\mathbb{Z}}f(2\delta k)=\frac{1}{\delta}\sqrt{\frac{\pi}{2}}\sum\limits_{m\in\mathbb{Z}}\widehat{f}(\frac{\pi m}{\delta}).
\end{equation}
To prove this equality one should calculate the Fourier coefficients for the periodic function $\sum\limits_{k\in\mathbb{Z}}f(x+2\delta k)$, construct the Fourier series and then set $x=0$.

Let us introduce the functions $g(x,\,p)=W(x)e^{-ipx}$ and $h(x,\,p)=W(x+\delta)e^{-ip(x+\delta)}$, which according to \eqref{tilde_W1_W2} give us
\begin{equation}
\label{g_h_func}
\sum\limits_{k}g(2k\delta,\,p)=\widetilde{W}_1(p),\quad \sum\limits_{k\in\mathbb{Z}}h(2k\delta,\,p)=\widetilde{W}_2(p).
\end{equation}

The Fourier transform for the initial function is $\widehat{W}(p)=F[W(x)](p)=\widehat{A}(p-k_0)$ where $\widehat{A}(p)=F[A(x)](p)$.
The corresponding Fourier transforms for the functions $g(x,\,p)$ and $h(x,\,p)$ are
$$
\widehat{g}(\omega,\,p)=F[g(x,\,p)](\omega)=\widehat{A}(\omega+p-k_0),\quad
\widehat{h}(\omega,\,p)=F[h(x,\,p)](\omega)=
e^{i\omega\delta}\widehat{A}(p+\omega-k_0).
$$
For each of these function we can apply the formula \eqref{Poisson_x_zero}. Using \eqref{g_h_func}, we have
$$
\widetilde{W}_1(p)=\frac{1}{\delta}\sqrt{\frac{\pi}{2}}\sum\limits_{m}\widehat{A}(\frac{\pi m}{\delta}+p-k_0),\quad
\widetilde{W}_2(p)=\frac{1}{\delta}\sqrt{\frac{\pi}{2}}\sum\limits_{m}(-1)^m\widehat{A}(\frac{\pi m}{\delta}+p-k_0).
$$

From these equalities we have
$$
\widetilde{W}_1(p)\pm \widetilde{W}_2(p)=\frac{1}{\delta}\sqrt{\frac{\pi}{2}}\sum\limits_{m}(1\pm(-1)^m)\widehat{A}(\frac{\pi m}{\delta}+p-k_0)
$$
Thus for the sum of functions $\widetilde{W}_1(p)+ \widetilde{W}_2(p)$ the non-zero terms in the right-hand side are only for the $m=2q$. This gives
$$
\widetilde{W}_1(p)+ \widetilde{W}_2(p)-\frac{\sqrt{2\pi}}{\delta}\widehat{A}(p-k_0)=\frac{\sqrt{2\pi}}{\delta}\sum\limits_{q\not=0}\widehat{A}(\frac{2\pi q}{\delta}+p-k_0).
$$
For $p\in[-\pi/(2\delta),\,\pi/(2\delta)]$ and $k_0\in[0,\,\pi/\delta]$ the argument $\frac{2\pi q}{\delta}+p-k_0\not=0$ for $q\not=0$. Therefore, after integration $n$ times by parts in the Fourier integral, we arrive to the following estimation
$$
\left|\widehat{A}\left(\frac{2\pi q}{\delta}+p-k_0\right)\right|\le  \frac{M_n}{\sqrt{2\pi}}\frac{\delta^n}{|2\pi q+\delta(p-k_0)|^n}\le \frac{M_n}{\sqrt{2\pi}}\frac{\delta^n}{|2\pi q-3\pi/2|^n},\quad 
M_n=\int\limits_{\mathbb{R}}|A^{(n)}(x)|dx.
$$
Thus we have
$$
|\widetilde{W}_1(p)+ \widetilde{W}_2(p)-\frac{\sqrt{2\pi}}{\delta}\widehat{A}(p-k_0)|\le M_n \delta^{n-1}\sum\limits_{q\not=0}\frac{1}{|2\pi q-3\pi/2|^n}.
$$
Because $A(\xi)$ belongs to the Schwartz space, we can take $n$ arbitrary large in this estimation. Because the series on the right-hand side converges absolutely, we have
\begin{equation}
\label{tilde_W1+W2}
\widetilde{W}_1(p)+ \widetilde{W}_2(p)-\frac{\sqrt{2\pi}}{\delta}\widehat{A}(p-k_0)=O(\delta^\infty).
\end{equation}

By the similar reasoning we have
\begin{equation}
\label{tilde_W1-W2}
\widetilde{W}_1(p)- \widetilde{W}_2(p)-\frac{\sqrt{2\pi}}{\delta}\widehat{A}(\frac{\pi}{\delta}+ p-k_0)=O(\delta^\infty).
\end{equation}
Formulas \eqref{tilde_W1+W2}, \eqref{tilde_W1-W2} are equivalent to the \eqref{tilde_W_1_Poisson} and \eqref{tilde_W_2_Poisson}.
\end{proof}

\subsection{Cauchy problem for the continuous equations.}

Before formulation of the Cauchy problem we want to introduce the operator $P=-i\mu\partial/\partial x$. The presence of the small parameter $\mu$ in this operator reflects the situation of localization of the initial functions (\ref{init_vec}) with the same parameter.  We have the following equality
$$
hD=-ih\frac{\partial}{\partial x}=\delta P,\quad  \delta=\frac{h}{\mu}.
$$ 

After that the Cauchy problem for the system of equations (\ref{main_sys_cont}), (\ref{main_operator}) becomes
\begin{equation}
\label{main_eq_Cauchy}
-h^2\frac{\partial^2}{\partial t^2}U(x,\,t)=2\Gamma L(\delta P)U(x,\,t),\quad U|_{t=0}=U^0\left(\frac{x}{\mu}\right),\,\frac{\partial}{\partial t}U\Bigr|_{t=0}=U^1\left(\frac{x}{\mu}\right).
\end{equation}
We start analysis of the equation \eqref{main_eq_Cauchy} with the symmetrization procedure. Let us denote the matrix $H(\delta p)=2\Gamma^{1/2}L(\delta p)\Gamma^{1/2}$, it is the self-adjoined and positive definite. Let $f_{1,\,2}(\delta p)$ be its the eigenvectors and $\omega_{1,\,2}^2(\delta p)$ be its eigenvalues.

Let us denote the matrices (projectors onto corresponding eigenspaces)
$$
\mathcal{P}_{1}(\delta p)=\frac{1}{|f_1(\delta p)|^2} f_1(\delta p)f_1^T(\delta p),\quad \mathcal{P}_{2}(\delta p)=\frac{1}{|f_2(\delta p)|^2} f_2(\delta p)f_2^T(\delta p).
$$

\begin{theorem}
\label{thm_general}

Let  the vectors $\widetilde{U}^{0,\,1}(p)$ be the semi-discrete Fourier transform of the \eqref{init_vec}, then the solution of the problem (\ref{main_eq_Cauchy}) has the following form 
\begin{equation}
\label{main_sys_Cauchy_sol}
U(x,\,t)=\sum\limits_{k=1,\,2}\frac{\delta}{\pi}\int\limits_{-\pi/(2\delta)}^{\pi/(2\delta)} \Gamma^{1/2}\mathcal{P}_{k}(\delta p)\Gamma^{-1/2}\left(\widetilde{U}^0(p)\cos(\frac{t}{h}\omega_k(\delta p))+\widetilde{U}^1(p)\frac{h}{\omega_k(\delta p)}\sin(\frac{t}{h}\omega_k(\delta p))\right)e^{\frac{i}{\mu}px}dp.
\end{equation}
\end{theorem}

\begin{proof}
Following definition \eqref{p_semi_Fourier} of the action of the operator $L(\delta P)$, we can rewrite the Cauchy problem  (\ref{main_eq_Cauchy}) in terms of the semi-discrete Fourier transform
$$
-h^2\frac{\partial^2}{\partial t^2}\widetilde{U}(p,\,t)=2\Gamma L(\delta p)\widetilde{U}(p,\,t),\quad \widetilde{U}|_{t=0}=\widetilde{U}^0(p),\,\frac{\partial}{\partial t}\widetilde{U}|_{t=0}=\widetilde{U}^1(p).
$$

Let us make the following change of the unknown function $\widetilde{V}(p,\,t)=\Gamma^{-1/2}\widetilde{U}(p,\,t)$. This function satisfies the problem
$$
-h^2\frac{\partial^2}{\partial t^2}\widetilde{V}(p,\,t)=H(\delta p)\widetilde{V}(p,\,t),\quad \widetilde{V}|_{t=0}=\Gamma^{-1/2}\widetilde{U}^0(p),\,\frac{\partial}{\partial t}\widetilde{V}|_{t=0}=\Gamma^{-1/2}\widetilde{U}^1(p).
$$
The general solution of this problem is given by the following formula
$$
\widetilde{V}(p,\,t)=\sum\limits_{k=1,\,2}\mathcal{P}_{k}(\delta p)\left(\Gamma^{-1/2}\widetilde{U}^0(p) \cos(\frac{t}{h}\omega_{k}(\delta p))+\Gamma^{-1/2}\widetilde{U}^1(p)\frac{h}{\omega_k(\delta p)}\sin(\frac{t}{h}\omega_2(\delta p))\right).
$$
Passing to the vector $\widetilde{U}(p,\,t)=\Gamma^{1/2}\widetilde{V}(p,\,t)$ and then applying the inverse semi-discrete Fourier transform lead to the statement of the theorem \ref{thm_general} and to the formula (\ref{main_sys_Cauchy_sol}).

\end{proof}

The formula (\ref{main_sys_Cauchy_sol}) gives the general form of the solution for the given Cauchy problem (\ref{main_eq_Cauchy}). It provides the decomposition of the solution into two modes: acoustical (eigenvalue $\omega_1(\delta p)$) and optical (eigenvalue $\omega_2(\delta p)$).  This formula depends on the parameter $\delta=h/\mu$ and we separate two cases: the case of the wide perturbation, $\delta\ll 1$ and the case of the narrow perturbation when $\delta<1$ but not small. In both these cases we simplify the integral (\ref{main_sys_Cauchy_sol}) and present it with the help of the Airy function and its derivative.

\section{Sufficiently wide perturbation.}
\label{sec_small_delta}

In this section we  provide an asymptotic analysis  of the solution (\ref{main_sys_Cauchy_sol}) when $\delta\ll 1$, which  means that the area of the initial perturbation $\mu$ is much larger than the lattice step $h$: $h\ll \mu$. It turns out that this condition is insufficient to determine the exact wave profile. For example, if $h=O(\mu^{3/2})$, than $\delta=\mu^{1/3}\ll 1$ and the dispersion effects  play the important role. On the other hand, if $h=O(\mu^{3/2+\alpha})$, $\alpha>0$ the dispersion effects are negligible. We are mostly interested in the dispersion case $h=O(\mu^{3/2})$.  We will also consider separately two cases: in-phase-oscillations,  when $k_0\approx 0$, and out-of-phase oscillations, when $k_0\approx \pi/\delta$.

Under the presented assumption we can simplify the representation of the solution \eqref{main_sys_Cauchy_sol}.
First of all we can simplify the matrices $\mathcal{P}_k(\delta p)$ and functions $\omega_k(\delta p)$. We have the following expansion for the matrices
\begin{equation}
\label{P_matr_long_wave}
\Gamma^{1/2}\mathcal{P}_1(\delta p)\Gamma^{-1/2}=\frac{1}{\gamma_1+\gamma_2}\begin{pmatrix}
\gamma_2  & \gamma_1\\
\gamma_2 & \gamma_1
\end{pmatrix}+O(\delta^2),\quad
\Gamma^{1/2}\mathcal{P}_2(\delta p)\Gamma^{-1/2}=\frac{1}{\gamma_1+\gamma_2}\begin{pmatrix}
\gamma_1  & -\gamma_1\\
-\gamma_2 & \gamma_2
\end{pmatrix}+O(\delta^2).
\end{equation}
Now let us present the expansion of the functions $\omega_k(\delta p)$. Let us denote
\begin{equation}
\label{omega_1_expans}
c_0=\sqrt{\frac{2\gamma_1\gamma_2}{\gamma_1+\gamma_2}},\quad q=c_0\frac{\gamma_1^2-\gamma_1\gamma_2+\gamma_2^2}{2(\gamma_1+\gamma_2)^{2}},
\end{equation}
then we have the following expansions
\begin{equation}
\label{omega_long_wave}
\omega_1(\delta p)=c_0 \delta |p|-\delta^3 \frac{1}{3} q  p^2|p|+O(\delta^5),\quad
\omega_2(\delta p)=\sqrt{2(\gamma_1+\gamma_2)}-\delta^2 \frac{c_0^2}{2\sqrt{2}\sqrt{\gamma_1+\gamma_2}} p^2+O(\delta^4).
\end{equation}

Following Lemma \ref{lm1_Poisson} we can simplify the initial conditions and replace the semi-discrete Fourier transform by the continuous Fourier transform
\begin{equation}
\label{init_con_long_wave}
\widetilde{U}^{0,\,1}(p)=\frac{1}{\delta}\sqrt{\frac{\pi}{2}}\widehat{A}^{0,\,1}(p-k_0)\begin{pmatrix}
1\\
1
\end{pmatrix}
+\frac{1}{\delta}\sqrt{\frac{\pi}{2}}\widehat{A}^{0,\,1}(\frac{\pi}{\delta}+p-k_0)\begin{pmatrix}
1\\
-1
\end{pmatrix}.
\end{equation}
From \eqref{P_matr_long_wave} we see, that the vector $(1,\,1)^T$ is the eigenvector of $\Gamma^{1/2}\mathcal{P}_1(0)\Gamma^{-1/2}$ and  also is in the kernel of $\Gamma^{1/2}\mathcal{P}_2(0)\Gamma^{-1/2}$. We say that this vector  corresponds to the acoustical mode. The vector $(1,\,-1)^T$ is not the eigenvector of $\Gamma^{1/2}\mathcal{P}_2(0)\Gamma^{-1/2}$ (recall $\gamma_1\not=\gamma_2$) and does not belong to the kernel of $\Gamma^{1/2}\mathcal{P}_1(0)\Gamma^{-1/2}$. The kernel of this matrix is the vector $(\gamma_1,\,-\gamma_2)^T$ which is the out-of-phase vector with corrected amplitudes for different atoms corresponding to the center of masses of the atomic cell.

\begin{theorem}
\label{thm_small_disp}
Let us suppose that $h^2=O(\mu^3)$  and $k_0\ll 1$. Let us define the following function
\begin{equation}
\label{main_sol_small_delta_disp}
U_{as}(x,\,t)=\frac{1}{\sqrt{2\pi}}\int\limits_{\mathbb{R}}\left(\widehat{A}^0(p-k_0)\cos(\frac{t}{\mu}\Phi_3(p))+\widehat{A}^1(p-k_0)\frac{\mu}{c_0|p|}\sin( \frac{t}{\mu}\Phi_3(p) )\right)e^{\frac{i}{\mu}px}dp.
\end{equation}
where $\Phi_3(p)=c_0|p|-\frac{\delta^2}{3}q p^2|p|$.
We have the following relation for the solution  (\ref{main_sys_Cauchy_sol}) of the initial system
$$
U(x,\,t)=U_{as}(x,\,t)
\left(
\begin{array}{c}
1 \\
1
\end{array}
\right)(1+O(\mu)).
$$

The function $U_{as}(x,\,t)$ satisfies the following problem
\begin{gather}
\label{approx_eq_small_disp}
-h^2\frac{\partial^2}{\partial t^2}U_{as}(x,\,t)=\left(c_0^2 \delta^2 P^2-\frac{2}{3} c_0 q\delta^4 P^4+\frac{1}{9}q^2\delta^6 P^6\right) U_{as}(x,\,t),\\
\label{approx_eq_small_disp_init_cond}
U_{as}(x,\,0)=A^0(\frac{x}{\mu})e^{\frac{i}{\mu}k_0x},\,\frac{\partial}{\partial t}	U_{as}(x,\,0)=A^1(\frac{x}{\mu})e^{\frac{i}{\mu}k_0x}.
\end{gather}
Here operator $P=-i\mu \partial/\partial x$ and the coefficients $c_0$ and $q$ are defined in (\ref{omega_1_expans}).

\end{theorem}

\begin{proof} Let us notice,  if $h^2=O(\mu^3)$, then $\delta^2=O(\mu)$. On the other hand from the expansion \eqref{omega_long_wave} we have the $\omega_1(\delta p)/h=(c_0|p|-\frac{\delta^2}{3}q p^2|p|)/\mu+O(h^4/\mu^5)$ and the correction $O(h^4/\mu^5)=O(\mu)$. This gives us the correction of order $O(\mu)$ for the solution. 

We substitute the initial conditions \eqref{init_con_long_wave} into the formula for the solution \eqref{main_sys_Cauchy_sol}. Since $k_0\ll 1$,  then the argument $\eta=\pi/\delta+ p-k_0$ belongs to the interval $\eta\in[\pi/(2\delta)-k_0,\,3\pi/(2\delta)-k_0]$ and is never zero. We can estimate functions $\widehat{A}^{0,\,1}(\eta)=O(\delta^\infty)$ by the similar reasoning as in the Lemma \ref{lm1_Poisson} by integration by parts of the Fourier integral.

The vector $(1,\,1)^T$ belongs to the kernel of the matrix $\Gamma^{1/2}\mathcal{P}_2(0)\Gamma^{-1/2}$. Thus we have only the integral corresponding to the acoustical mode
$$
U(x,\,t)=\frac{1}{\sqrt{2\pi}}\int\limits_{-\pi/(2\delta)}^{\pi/(2\delta)} \left(\widehat{A}^0(p-k_0)\cos(\frac{t}{h}\omega_1(\delta p))+\widehat{A}^1(p-k_0)\frac{h}{\omega_1(\delta p)}\sin(\frac{t}{h}\omega_1(\delta p))\right)e^{\frac{i}{\mu}px}dp\left(
\begin{array}{c}
1 \\
1
\end{array}
\right).
$$
After that we pass to the integration by whole line, since $\delta\ll 1$ and $k_0$ is small.  Finally we use the expansion \eqref{omega_long_wave} for $\omega_1(\delta p)$ and truncate the small correction.

We can easily verify that the function satisfies the Cauchy problem \eqref{approx_eq_small_disp}, \eqref{approx_eq_small_disp_init_cond}. The verification of the initial condition is trivial and can be done by the direct computation. The form of the equation \eqref{approx_eq_small_disp} follows from the equality 
$$
(c_0\delta|p|-\frac{\delta^3}{3}q p^2|p|)^2=c_0^2\delta^2 p^2-\frac{2}{3}c_0q\delta^4 p^4+\frac{1}{9}q^2 \delta^6 p^6.
$$
Thus the square root of the symbol of the operator in the equation gives the exact terms in the expansion of the $\omega_1(\delta p)$.
\end{proof}

From the representation \eqref{init_con_long_wave} and the Theorem \ref{thm_small_disp} we can see, that we have two regimes, separated by the Nyquist frequency $k_N=\pi/(2\delta)$. If $k_0<k_N$,  then the initial perturbation excites only the acoustical mode. 

\begin{remark}
From the proof of the Theorem \ref{thm_small_disp}, we can see that the symbol of the approximate equation \eqref{approx_eq_small_disp} is positive and equals to zero in the points $p=0$ and $p=\pm (1/\delta)\sqrt{3c_0/q}$. This symbol is not the Taylor expansion of the function $\omega_1(\delta p)$, but it gives the well-posed approximate equation and moreover it is simpler than the Pade approximation and we also have the exact solution of this approximate equation.
\end{remark}

When the initial perturbation is in the form of the Gaussian exponential, the integral (\ref{main_sol_small_delta_disp}) can be analytically calculated and presented with the help of the Airy function \cite{Olver}
\begin{equation}
\label{Airy_func_def}
Ai(z)=\frac{1}{2\pi i}\int\limits_{\infty e^{-i\pi/3}}^{\infty e^{i\pi/3}}e^{\frac{u^3}{3}-zu}du.
\end{equation}
For simplicity we choose the initial velocity to be zero.

\begin{corollary}
\label{cor_Gauss_small_disp}
Let in the equation \eqref{main_sol_small_delta_disp} the functions be $\widehat{A}^0(\xi)=-e^{-\xi^2/2}$ and $\widehat{A}^1(\xi)=0$.
Let us introduce the following quantities
$$
A^+=\frac{x- c_0 t}{\mu},\,A^-=-\frac{x+ c_0 t}{\mu},\, \alpha=\delta^{2/3} \left(\frac{t q}{\mu}\right)^{1/3},
$$
then the integral (\ref{main_sol_small_delta_disp}) equals
\begin{equation}
\label{U_small_disp_Airy_summ}
U_{as}(x,\,t)=\sqrt{\frac{\pi}{2}}\frac{1}{\alpha}\sum\limits_{\pm}e^{\frac{2}{3\alpha^6}(\frac{1}{2}\mp i\alpha^3k_0)^3}e^{\frac{1}{2}\left(\frac{A^\pm}{\alpha^3}+k_0\right)}e^{\mp i\frac{2}{3}\alpha^3 k_0^3}Ai\left(\frac{A^\pm}{\alpha}+ \alpha^2 k_0^2+\frac{1}{\alpha^4}\left(\frac{1}{2}\mp i\alpha^3 k_0\right)^2\right).
\end{equation}
Moreover, if $k_0=0$, then we have
\begin{equation}
\label{U_small_disp_Airy_summ_k0_zero}
U_{as}(x,\,t)=\sqrt{\frac{\pi}{2}}\frac{1}{\alpha}e^{\frac{1}{12\alpha^6}}\left(e^{ \frac{x- c_0 t}{2\alpha^3 \mu}}Ai\left( \frac{x- c_0 t}{\mu \alpha}+\frac{1}{4\alpha^4}\right)+e^{- \frac{x+ c_0 t}{2\alpha^3 \mu}}Ai\left(- \frac{x+ c_0 t}{\mu \alpha}+\frac{1}{4\alpha^4}\right)\right).
\end{equation}

\end{corollary}

\begin{proof}
We follow ideas of \cite{Vallee}. Suppose $\alpha>0$, then we use the following representation
$$
\frac{1}{2\pi}\int\limits_{\mathbb{R}}e^{-\frac{(p-k_0)^2}{2}}e^{i A p}e^{i\frac{\alpha^3}{3}p^3}dp=\frac{1}{\alpha}e^{i(A k_0+\frac{\alpha^3}{3}k_0^3)}e^{\frac{2}{3}\frac{z^3}{\alpha^6}+(A+\alpha^3 k_0^2)\frac{z}{\alpha^3}}Ai\left(\frac{z}{\alpha^4}+\frac{A}{\alpha}+\alpha^2 k_0^2\right),
$$
where $z=1/2-i\alpha^3k_0$. This representation follows from the following sequence of changes of variables: $p=\eta+k_0$, $\eta=v- iz/\alpha^3$ and $u=\alpha v$, which give the integral of the Airy type over contour $Im\, u=1/2\alpha^2$. After the change of variables $w=-i u$, we have the integration over the contour $Re\, w=1/2\alpha^2$. After that, we use the fact that the function under the integral is the entire function and thus we can deform the contour of the integration to the contour of the Airy function, similar to \cite{Bleistein}. The simplification of the formula leads to \eqref{U_small_disp_Airy_summ}. In the case of the negative cubic part  $-\alpha^3/3p^3$ of the phase in the integral, the reasoning is the same except of the last step when we transform the integral to the integral over $Re\, w=-1/2\alpha^2$ with the change of variables $w=i u$.
\end{proof}

\begin{remark}
If in Theorem \ref{thm_small_disp} we suppose that $h^2=O(\mu^{3+\alpha})$ where $\alpha>0$, then the cubic correction to the phase $\Phi_3(p)/\mu$ will be of order $O(\mu^\alpha)$. In this case the function $U_{as}(x,\,t)$ satisfies the wave equation
$$
-h^2\frac{\partial^2}{\partial t^2}U_{as}(x,\,t)=c_0^2 \delta^2 P^2 U_{as}(x,\,t),\,
U_{as}(x,\,0)=A^0(\frac{x}{\mu})e^{\frac{i}{\mu}k_0x},\,\frac{\partial}{\partial t}	U_{as}(x,\,0)=A^1(\frac{x}{\mu})e^{\frac{i}{\mu}k_0x}.
$$
The solution of the system \eqref{main_sys_Cauchy_sol} is of the form $U(x,\,t)=U_{as}(x,\,t)(1,\,1)^T(1+O(\mu^\alpha))$.
\end{remark}

Now let us present the situation for the out-of-phase initial conditions.
\begin{theorem}
\label{thm_out_phase}
Let us suppose that $h^2=O(\mu^3)$ and $\pi/\delta-k_0=\omega_0>0$ where $\omega_0\ll 1$. Let us define the functions
\begin{gather}
\label{out_phase_as_ac}
U^{ac}_{as}(x,\,t)=\frac{\gamma_2-\gamma_1}{\gamma_1+\gamma_2}\frac{1}{\sqrt{2\pi}}\int\limits_{\mathbb{R}}\left(\widehat{A}^0(p-\omega_0)\cos(\frac{t}{\mu}\Phi_3(p))+\widehat{A}^1(p-\omega_0)\frac{\mu}{c_0|p|}\sin( \frac{t}{\mu}\Phi_3(p) )\right)e^{\frac{i}{\mu}px}dp,\\
\label{out_phase_as_opt}
U^{opt}_{as}(x,\,t)=\frac{2}{\gamma_1+\gamma_2}\frac{1}{\sqrt{2\pi}}\int\limits_{\mathbb{R}}\left(\widehat{A}^0(p-\omega_0)\cos(\frac{t}{h}\Phi_2(p))+h\frac{\widehat{A}^1(p-\omega_0)}{\sqrt{2(\gamma_1+\gamma_2)}}\sin( \frac{t}{h}\Phi_2(p) )\right)e^{\frac{i}{\mu}px}dp.
\end{gather}
where $\Phi_3(p)=c_0|p|-\frac{\delta^2}{3}q p^2|p|$ and $\Phi_2(p)=\sqrt{2(\gamma_1+\gamma_2)}-\delta^2\frac{c_0^2}{2\sqrt{2(\gamma_1+\gamma_2)}}p^2$.

Then the solution  (\ref{main_sys_Cauchy_sol}) of the initial system has the form
\begin{equation}
\label{modes_separation}
U(x,\,t)=U^{ac}_{as}(x,\,t)
\left(
\begin{array}{c}
1 \\
1
\end{array}
\right)(1+O(\mu))+
U^{opt}_{as}(x,\,t)
\left(
\begin{array}{c}
\gamma_1 \\
-\gamma_2
\end{array}
\right)(1+O(\sqrt{\mu})).
\end{equation}
\end{theorem}

\begin{proof} Proof is similar to the proof of the Theorem \ref{thm_small_disp}, and thus we will just indicate differences. The first difference is the order $O(\sqrt{\mu})$ for the optical mode. This comes from the expansion \eqref{omega_long_wave} of the $\omega_2(\delta p)$ and the correction order $O(\delta^4)$ in it. Thus the correction order for the $\omega_2(\delta p)/h$ is of order $O(h^3/\mu^4)=O(\sqrt{\mu})$, when  $h^2=O(\mu^3)$.

We substitute the initial conditions \eqref{init_con_long_wave} into the formula for the solution \eqref{main_sys_Cauchy_sol}. Since $\pi/\delta-k_0=\omega_0>0$ where $\omega_0\ll 1$ then the argument of the functions $\widehat{A}^{0,\,1}(p-k_0)$ is never is zero for $p\in[-\pi/(2\delta),\,\pi/(2\delta)]$. Therefore we can integrate the Fourier integral by parts. This leads to the $O(\delta^{\infty})$ correction. 

The vector $(1,\,-1)^T$ does not belong to the kernels of the matrices $\Gamma^{1/2}\mathcal{P}_{1,\,2}(0)\Gamma^{-1/2}$. The application of these matrices to this vector leads to the separation of the modes in \eqref{modes_separation}.  After that we pass to the integration by the whole line, since $\delta\ll 1$ and $\omega_0$ is small and after some calculations we arrive the the desired formula.
\end{proof}

We can see that in both cases of smooth or out-of-phase initial conditions, the asymptotic functions \eqref{out_phase_as_ac} and \eqref{U_small_disp_Airy_summ} have the same structure and the difference between them is only the multiplication constant $(\gamma_2-\gamma_1)/(\gamma_1+\gamma_2)$.

\begin{corollary}
\label{cor_opt_wave_eq}
Function \eqref{out_phase_as_opt} satisfies the equation
\begin{equation}
\label{opt_wave_eq}
-h^2 \frac{\partial^2}{\partial t^2} U_{as}^{opt}(x,\,t)=\left[2(\gamma_1+\gamma_2)-c_0^2\delta^2 P^2+\frac{\delta^4 c_0^4P^4}{8(\gamma_1+\gamma_2)}\right] U_{as}^{opt}(x,\,t),
\end{equation}
operator $P=-i\mu \partial/\partial x$ and the coefficient $c_0$ is defined in (\ref{omega_1_expans}).
\end{corollary}

\begin{proof}
This can be easily verified by the same method as for the acoustical mode. The form of the equation \eqref{opt_wave_eq} follows from the equality 
$$
\left(\sqrt{2(\gamma_1+\gamma_2)}-\delta^2\frac{c_0^2}{2\sqrt{2(\gamma_1+\gamma_2)}}p^2\right)^2=2(\gamma_1+\gamma_2)-c_0^2\delta^2p^2+\frac{\delta^4 c_0^4 p^4}{8(\gamma_1+\gamma_2)}.
$$
Thus the square root of the symbol of the operator in the equation \eqref{opt_wave_eq} gives the exact terms in the expansion of the $\omega_2(\delta p)$.

\end{proof}

\begin{remark}
The equation \eqref{opt_wave_eq} is not the wave equation type, since the derivative in the spatial variable $x$ has the opposite sign. 
This equation is well-posed since the symbol of the differential operator in spatial variable is non-negative.
\end{remark}

\begin{corollary}
\label{cor_gaus_exp_opt}
In the terms of the Theorem \ref{thm_out_phase}, let $A^0(\xi)=e^{-\xi^2/2}$ and $\widehat{A}^1(\xi)\equiv 0$, then for the optical wave mode \eqref{out_phase_as_opt} we have the following expression
\begin{equation}
\label{opt_Gauss}
U^{opt}_{as}(x,\,t)=\frac{\sqrt{\mu}}{\gamma_1+\gamma_2}\sum\limits_{\pm}\frac{e^{\pm i\frac{t}{h}\sqrt{2(\gamma_1+\gamma_2)}}}{(\mu^2+ 4\delta^2t^2 a_\mp^2)^{1/4}}e^{-\frac{(x+2\delta t a_\mp\omega_0)^2}{2(\mu^2+4\delta^2t^2a_\mp^2)}}e^{i\frac{\delta ta_\mp(\mu^2\omega_0^2-x^2)/\mu+\mu\omega_0x}{\mu^2+4\delta^2t^2a_\mp^2}}e^{i\frac{\arctan(2\delta ta_\mp/\mu)}{2}},
\end{equation}
where $a_\mp=\mp  c_0^2 /(2\sqrt{2(\gamma_1+\gamma_2)})$. Moreover, when $\omega_0=0$ we have
\begin{equation}
\label{opt_Gauss_omega_0}
U^{opt}_{as}(x,\,t)=\frac{\sqrt{\mu}}{\gamma_1+\gamma_2}\sum\limits_{\pm}\frac{e^{\pm i\frac{t}{h}\sqrt{2(\gamma_1+\gamma_2)}}}{(\mu^2+ 4\delta^2t^2 a_\mp^2)^{1/4}}e^{-\frac{x^2}{2(\mu^2+4\delta^2t^2a_\mp^2)}}e^{-i\frac{tx^2\delta a_\mp}{\mu(\mu^2+4 \delta^2 t^2a_\mp^2)}}e^{i\frac{\arctan(2\delta ta_\mp/\mu)}{2}},
\end{equation}

\end{corollary}

\begin{proof}
This follows from the equality for the inverse Fourier transform of the Gaussian wave packet
$$
\frac{1}{\sqrt{2\pi}}\int\limits_{\mathbb{R}}e^{-\frac{(p-\omega_0)^2}{2}}e^{i a p^2}e^{i p \xi}dp=\frac{1}{(1+ 4 a^2)^{1/4}}e^{-\frac{(\xi+2a\omega_0)^2}{2(1+4a^2)}}e^{i\frac{a(\omega_0^2-\xi^2)+\omega_0\xi}{1+4a^2}}e^{i\frac{\arctan(2a)}{2}}.
$$
\end{proof}

\begin{remark}
\label{rm_Gouy} Here we want to draw an attention to the phase correction $\arctan(2\delta ta_\mp/\mu)/2$ in \eqref{opt_Gauss}, \eqref{opt_Gauss_omega_0}. This correction by the form reminds the Gouy phase shift \cite{Boyd, Feng}, but in the time variable. Thus we can define  the Rayleigh time as $t_R=\mu\sqrt{2(\gamma_1+\gamma_2)} /(\delta c_0^2)$. Note that this Gouy phase shift depends on parameter $\delta$ and for the long wave approximation it is negligible, starting to play role only for dispersive model.
\end{remark}

\section{Case of the sufficiently narrow perturbation. Acoustical mode.}
\label{sec_strong_disp}
This is the boundary case when the Kotel'nikov-Whittaker-Shannon interpolation of the initial data gives the small correction. In this case $\delta<1$, but not very small.  We treat the initial conditions fast-changing and thus we consider the case without any oscillatory part in the form of the initial conditions \eqref{Gauss_pack_lattice}, meaning that $k_0=0$. For simplicity we  also consider the situation when the initial velocity is zero, $\widetilde{U}^1(p)\equiv 0$.

We can apply the Lemma \ref{lm1_Poisson}, and following our assumptions, we arrive that in this setup we have the initial  perturbation of the form
\begin{equation}
\label{short_acoust_init}
\widetilde{U}^0(p)\approx\widehat{A}(p)\
\begin{pmatrix}
1\\
1
\end{pmatrix},
\end{equation}
where $\widehat{A}(p)$ is the Fourier transform of the initial function. The numerical simulation shows that in this case the main wave is the wave corresponding to the acoustical branch, while amplitude of the optical mode wave is negligible compare to the acoustical. Further we will consider the acoustical mode.

According to the general form of the solution \eqref{main_sys_Cauchy_sol}, the acoustical wave  is described by the following integral
\begin{equation}
\label{I1_strong}
I^{ac}(x,\,t)=\frac{1}{\pi}\int\limits_{-\pi/(2\delta)}^{\pi/(2\delta)}\Gamma^{1/2}\mathcal{P}_1(\delta p)\Gamma^{-1/2}\widetilde{U}^0(p)\cos(\frac{i}{h}\omega_1(\delta p)t)e^{\frac{i}{\mu}px}dp.
\end{equation}

This integral can be rewritten using the representation \eqref{short_acoust_init} and according to the Statement \ref{st_Kot_correc}, we also can pass to the integration over the whole line. But we present here more general approach for the asymptotic calculation of this integral, and thus we want to preserve initial form of the integral.

We provide the asymptotic representation of the integral \eqref{I1_strong} via  Airy function and its derivative near some small vicinity of the wave front, see \cite{sergeev_asymptotic_2019}. Then, following the ideas of \cite{anikin_uniform_2019}, we construct the so-called uniform representation of the asymptotic solution via Airy functions in the much larger area. 

The principal difference from the previous case is that the representation via Airy functions can be given for broader forms of the initial conditions, not only for the Gaussian exponential.

We start from the representation of the integral $I^{ac}(x,\,t)$ via Airy functions near the wave front. 
Let us write down the expansion of the function $\omega_1(\delta p)$ for small $|p|\ll 1$
$$
\omega_1(\delta p)=c_0\delta  |p|-q\delta^3 \frac{p^2 |p|}{3}+O(|p|^5),\quad
c_0=\sqrt{\frac{2\gamma_1\gamma_2}{\gamma_1+\gamma_2}},\quad q=c_0\frac{\gamma_1^2-\gamma_1\gamma_2+\gamma_2^2}{2(\gamma_1+\gamma_2)^{2}}.
$$

Similar to \cite{dobrokhotov_maslovs_2014} we can introduce the following functions
\begin{equation}
\label{I12_cor_V1_V2_strong_disp}
\widetilde{U}^0_{1}(z)=\frac{1}{2}(\widetilde{U}^0(\sqrt{z})+\widetilde{U}^0(-\sqrt{z})),\quad 
\widetilde{U}^0_{2}(z)=\frac{1}{2\sqrt{z}}(\widetilde{U}^0(\sqrt{z})-\widetilde{U}^0(-\sqrt{z})),\quad z\ge 0.
\end{equation}

Near the wavefront the asymptotic representation is more or less simple, and it is given by the following Statement.

\begin{statement}
\label{st_I11_strong_disp}

In the $O(\mu^{2/3})$-vicinity of the wave front $x=c_0t$,   we have the following
\begin{gather}
\label{I12_Airy_strong}
Re\,I^{ac}(x,\,t)=\Gamma^{1/2}\mathcal{P}_1(0)\Gamma^{-1/2} Re\Biggl(\left(\frac{\mu\delta}{qt}\right)^{1/3}\widetilde{U}^0_{1}\left(-\frac{x-c_0t}{q\delta^2 t}\right) Ai\left(\frac{x-c_0t}{\mu^{2/3}\delta^{2/3}(qt)^{1/3}}\right)-\\
\nonumber
-i\frac{1}{\delta^{1/3}}\left(\frac{\mu}{qt}\right)^{2/3}\widetilde{U}^0_{2}\left(-\frac{x-c_0t}{q\delta^2 t}\right)Ai'\left(\frac{x-c_0t}{\mu^{2/3}\delta^{2/3}(qt)^{1/3}}\right)\Biggr)(1+O(\mu^{2/3})).
\end{gather}

In the $O(\mu^{2/3})$-vicinity of the wave front $x=-c_0t$ we have the similar representation with the help of the Airy function
\begin{gather}
\nonumber
Re\, I^{ac}(x,\,t)=\Gamma^{1/2}\mathcal{P}_1(0)\Gamma^{-1/2} Re\Biggl(\left(\frac{\mu \delta }{qt}\right)^{1/3}\widetilde{U}^0_{1}\left(\frac{x+c_0t}{q\delta^2  t}\right) Ai\left(-\frac{x+c_0t}{\mu^{2/3}\delta^{2/3}(qt)^{1/3}}\right)+\\
\label{I11_Airy_strong_disp}
+i\frac{1}{\delta^{1/3}}\left(\frac{\mu}{qt}\right)^{2/3}\widetilde{U}^0_{2}\left(\frac{x+c_0t}{q\delta^2 t}\right)Ai'\left(-\frac{x+c_0t}{\mu^{2/3}\delta^{2/3} (qt)^{1/3}}\right)\Biggr)(1+O(\mu^{2/3})).
\end{gather}
\end{statement}

When $|x|>c_0t$ the integral $I^{ac}(x,\,t)$ is of order $O(\mu^{\infty})$. Since Airy function and its derivative are exponentially decaying for $|x|>c_0 t$, we need to smoothly continue the functions $\widetilde{U}^0_{1,\,2}(z)$ for $z\le 0$.  We use the three-points approximation similar to \cite{dobrokhotov_maslovs_2014, anikin_uniform_2019}
$$
\widetilde{U}^{0-}_{1}(z)=3\widetilde{U}^0_{1}(0)-3\widetilde{U}^0_{1}(-z)+\widetilde{U}^0_{1}(-2z),\quad
\widetilde{U}^{0-}_{2}(z)=3\widetilde{U}^0_{2}(0)-3\widetilde{U}^0_{2}(-z)+\widetilde{U}^0_{2}(-2z).
$$
This allows us to smoothly extend the functions (\ref{I12_Airy_strong}) and (\ref{I11_Airy_strong_disp}) for $|x|>ct$ with the order $O(\mu^{\infty})$.

Using the representation \eqref{short_acoust_init} and the fact that the vector $(1,\,1)^T$ is the eigenvector of the matrix $\Gamma^{1/2}\mathcal{P}_1(0)\Gamma^{-1/2}$, we arrive to the following corollary.
\begin{corollary}
\label{cor_I11_strong_disp}
Using the representation \eqref{short_acoust_init}, in Statement \ref{st_I11_strong_disp}, we have the following asymptotic near the wave front.

In the $O(\mu^{2/3})$-vicinity of the wave front $x=c_0t$,   we have the following
\begin{gather}
\label{I12_Airy_strong_A}
Re\,I^{ac}(x,\,t)=\begin{pmatrix}
1\\
1
\end{pmatrix} 
Re\Biggl(\left(\frac{\mu\delta}{qt}\right)^{1/3}\widehat{A}^0_{1}\left(-\frac{x-c_0t}{q\delta^2 t}\right) Ai\left(\frac{x-c_0t}{\mu^{2/3}\delta^{2/3}(qt)^{1/3}}\right)-\\
\nonumber
-i\frac{1}{\delta^{1/3}}\left(\frac{\mu}{qt}\right)^{2/3}\widehat{A}^0_{2}\left(-\frac{x-c_0t}{q\delta^2 t}\right)Ai'\left(\frac{x-c_0t}{\mu^{2/3}\delta^{2/3}(qt)^{1/3}}\right)\Biggr)(1+O(\mu^{2/3})).
\end{gather}

In the $O(\mu^{2/3})$-vicinity of the wave front $x=-c_0t$ we have the similar representation with the help of the Airy function
\begin{gather}
\nonumber
Re\, I^{ac}(x,\,t)=\begin{pmatrix}
1\\
1
\end{pmatrix}
 Re\Biggl(\left(\frac{\mu \delta }{qt}\right)^{1/3}\widehat{A}^0_{1}\left(\frac{x+c_0t}{q\delta^2  t}\right) Ai\left(-\frac{x+c_0t}{\mu^{2/3}\delta^{2/3}(qt)^{1/3}}\right)+\\
\label{I11_Airy_strong_disp_A}
+i\frac{1}{\delta^{1/3}}\left(\frac{\mu}{qt}\right)^{2/3}\widehat{A}^0_{2}\left(\frac{x+c_0t}{q\delta^2 t}\right)Ai'\left(-\frac{x+c_0t}{\mu^{2/3}\delta^{2/3} (qt)^{1/3}}\right)\Biggr)(1+O(\mu^{2/3})).
\end{gather}
Functions $\widehat{A}^{0}_{1,2}(p)$ are defined similarly to \eqref{I12_cor_V1_V2_strong_disp}. The continuation process for the negative values is also similar.
\end{corollary}

One can replace the representation near the wave front, given in Statement \ref{st_I11_strong_disp},  by the other representation, valid in more larger interval. This new representation is based on the fact, that the Airy function can be approximated by the WKB form in the area of oscillations. Thus we need to construct the uniform argument for the Airy function in such a way, it connects two representations. It cannot be done exactly, but with sufficient error, which does not reduce the asymptotic correction. This  uniform representation is given in the following Theorem.

\begin{theorem}
\label{thm_acoustic_I12_strong_disp}
Let the function $p_1=p_1(x,\,t)\ge 0$ be the positive solution of the equation $x+\omega_1'(\delta p_1)t=0$ when $-c_0t<x\le 0$. Let us define $p_2=p_2(x,\,t)\ge 0$ as the positive solution of the equation $x-\omega_1'(\delta p_2)t=0$ when $0\le x<c_0t$.

Let us introduce the  functions
$$
S_1(p_1)=p_1 x+\frac{1}{\delta}\omega_1(\delta p_1)t,\quad S_2(p_2)=-(p_2 x-\frac{1}{\delta}\omega_1(\delta p_2)t).
$$

Then for some $\varepsilon>0$ when $x\in(-c_0t-\varepsilon,\,0]$, we have the following representation 
\begin{gather}
\label{I1_unif_left_wave_strong_disp}
Re\,I^{ac}(x,\,t)=\sqrt{2\delta \mu}\frac{\Gamma^{1/2}\mathcal{P}_1(\delta p_1)\Gamma^{-1/2}}{\sqrt{|\omega_1''(\delta p_1)|}\sqrt{t}} \Biggl(\widetilde{U}_1^0(p_1^2)\left(\frac{3}{2}\frac{S_1(p_1)}{\mu}\right)^{1/6}Ai\left(-\left(\frac{3}{2}\frac{S_1(p_1)}{\mu}\right)^{2/3}\right)+\\
\nonumber
+ip_1\widetilde{U}_2^0(p_1^2)\left(\frac{3}{2}\frac{S_1(p_1)}{\mu}\right)^{-1/6}Ai'\left(-\left(\frac{3}{2}\frac{S_1(p_1)}{\mu}\right)^{2/3}\right)\Biggr)(1+O(\mu^{2/3})).
\end{gather}

When $x\in [0,\,c_0t+\varepsilon)$, then we have the following representation
\begin{gather}
\label{I1_unif_right_wave_strong_disp}
Re\,I^{ac}(x,\,t)=\sqrt{2\delta \mu}\frac{\Gamma^{1/2}\mathcal{P}_1(\delta p_2)\Gamma^{-1/2}}{\sqrt{|\omega_1''(\delta p_2)|}\sqrt{t}} \Biggl(\widetilde{U}_1^0(p_2^2)\left(\frac{3}{2}\frac{S_2(p_2)}{\mu}\right)^{1/6}Ai\left(-\left(\frac{3}{2}\frac{S_2(p_2)}{\mu}\right)^{2/3}\right)-\\
\nonumber
-ip_2\widetilde{U}_2^0(p_2^2)\left(\frac{3}{2}\frac{S_2(p_2)}{\mu}\right)^{-1/6}Ai'\left(-\left(\frac{3}{2}\frac{S_2(p_2)}{\mu}\right)^{2/3}\right)\Biggr)(1+O(\mu^{2/3})).
\end{gather}

\end{theorem}
Proof of the Statement \ref{st_I11_strong_disp} and of the Theorem \ref{thm_acoustic_I12_strong_disp} is given in the Appendix \ref{app_stron_disp}.

Note, that if in Statement \ref{st_I11_strong_disp} and Theorem  \ref{thm_acoustic_I12_strong_disp}, the initial vector-function $\widetilde{U}^0(p)$ is even, then the asymptotic representation can be significantly simplified. Moreover in this case, there is no any derivative of the Airy function in the formulas.



\section{Example for perfect lattice.}
\label{sec_exmpl}
In this section we are providing  examples for the perfect diatomic one-dimensional lattice. We took the parameters of this lattice to be the same as for the NaCl lattice.

Parameters $\gamma_{1,2}$ depend only on the mass ratio. We have $m_{Cl}=35.446$ and $m_{Na}=22.989$ in standard atomic weights, thus
$$
\gamma_1=\gamma_{Cl}=\frac{m_{Cl}+m_{Na}}{2 m_{Cl}}\approx 0.82,\quad \gamma_2=\gamma_{Na}=\frac{m_{Cl}+m_{Na}}{2 m_{Cl}}\approx 1.27.
$$
Here for the normalization we chose the average mass $m_0=\sqrt{2 m_{Cl}^2 m_{Na}^2/(m_{Cl}^2 +m_{Na}^2)}$ and following \eqref{omega_1_expans} and we have the dimensionless parameters of the acoustical branch: the velocity  $c_0=1$ and the dispersion parameter $q\approx 0.14$.

 The  distance between two nearest atoms in NaCl is $d=2.82$ \AA\,  and  the length of the lattice we chose as $L=10^4$ \AA. The point of the initial perturbation is the origin, and it is in the middle of the lattice.
We normalize the length $L$ by dimensionless length 2 and the lattice step is $h=0.000282$. In this case the matrix of the original system has the dimensions $7093\times 7093$ and the reference numerical solution is obtained by direct computation of the eigenvalues and eigenvectors of the matrix and further standard determination of the dynamics of the coefficients of the expansion.

Here we assume that the amplitude of the initial conditions is in the Gaussian form.
We start with the case of sufficiently large area of the initial perturbation, meaning that $\delta\ll 1$ . We choose $\mu=h^{2/3}/2\approx 0.00215$ and $\delta=h/\mu\approx 0.1311$. The Nyquist frequency is $k_N=\pi/(2\delta)\approx 11.9768$.

\begin{figure}[!h]
\begin{center}
\includegraphics[scale=0.6]{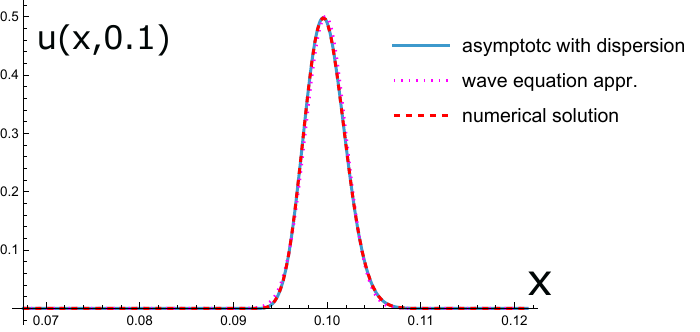}
\includegraphics[scale=0.6]{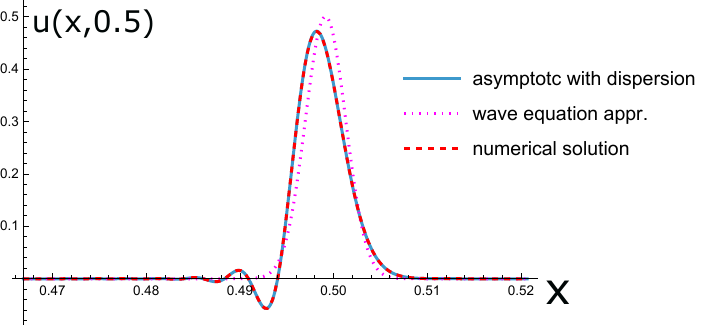}
\end{center}
\caption{Graphs of the amplitude of the oscillations of the atoms. The solid line represents the asymptotic formula (\ref{U_small_disp_Airy_summ}), the dashed line represents the numerical solution and the dotted line represents the solution of the wave equation. We chose time moments $t=0.1$ (on the left) and $t=0.5$ (on the right). \label{pic_smooth_ac_disp}}
\end{figure}

We start the example from the case of the smooth in-phase initial condition. On the figure \ref{pic_smooth_ac_disp} we plotted the numerical solution of the system \eqref{osc_lattice_dimens}, asymptotic solution \eqref{U_small_disp_Airy_summ_k0_zero} and the solution of the wave equation to compare with for the different time moments. We chose the wave which propagates to the right. From this comparison we clearly see that the dispersion effects are playing a significant role during the wave propagation and they change the wave profile.

The next example corresponds to the out-of-phase initial condition, when the initial frequency $k_0=2k_N\approx 23.9536$. Interestingly enough, we want to notice that this case is sensitive to the lattice numeration, because $e^{i k_0 h n/\mu}=(-1)^n$, where $n$ is the number of lattice point. Contrary, the case of the smooth initial condition does not depend on the number of each point in the lattice. So, we fix the lattice in a way, that on the left and on the right, with respect to the origin, we have the even number of points.

In this case we have excitation of the waves over both modes, acoustical and optical. The acoustical wave differs from the case of smooth initial conditions only by multiplier $(\gamma_2-\gamma_1)/(\gamma_1+\gamma_2)$. Therefore, we can also represent this wave using Airy representation \eqref{U_small_disp_Airy_summ_k0_zero}, adjusting the amplitude. 

The wave which corresponds to the optical mode is fast oscillating function, and the function \eqref{opt_Gauss_omega_0} multiplied by the vector $(\gamma_1,\,-\gamma_2)^T$, defines the envelope curves for this solution. In order to obtain the solution one has to restrict each curve onto sublattices, each corresponding to the types of atoms.

\begin{figure}[!h]
\begin{center}
\includegraphics[scale=1]{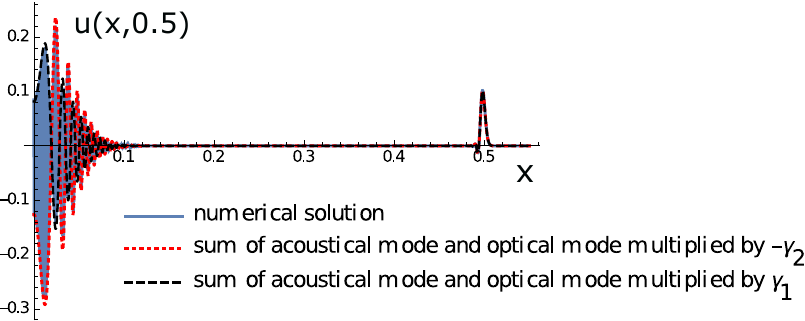}
\end{center}
\caption{Graphs of the amplitudes corresponding to the acoustical \eqref{out_phase_as_ac} and optical \eqref{opt_Gauss_omega_0} waves. The continuous line corresponds to the numerical solution, the dashed line corresponds to the sum of acoustical mode and the optical mode multiplied by $-\gamma_2$. The dotted line is the sum of the acoustical wave and optical wave multiplied by $\gamma_1$. Time moment is $t=0.5$. \label{pic_out_phase_sum}}
\end{figure}

\begin{figure}[!h]
\begin{center}
\includegraphics[scale=0.6]{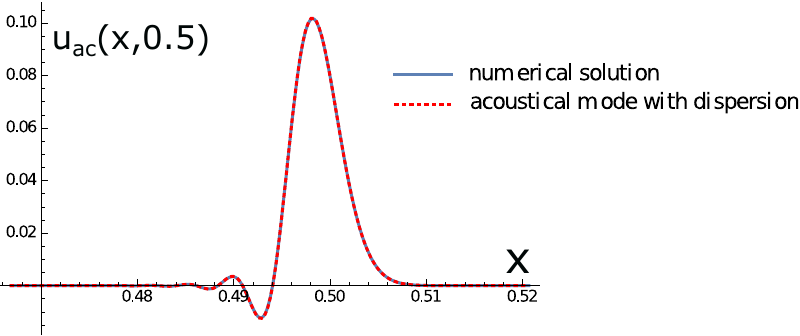}
\includegraphics[scale=0.6]{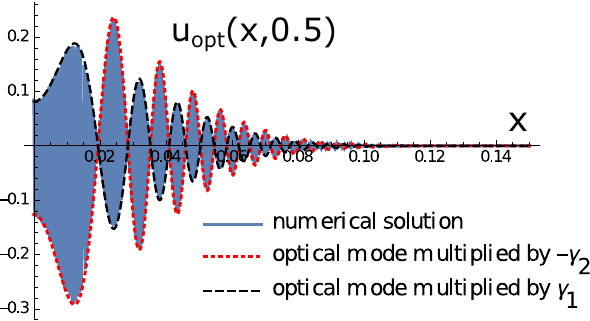}
\end{center}
\caption{Zoom of the amplitudes from the pic. \ref{pic_out_phase_sum}. On the left we have the acoustical wave, which is the multiplication of \eqref{U_small_disp_Airy_summ_k0_zero} by $(\gamma_2-\gamma_1)/(\gamma_1+\gamma_2)$. On the right is the optical mode. We plotted two envelope curves of the numerical solution, given by  \eqref{opt_Gauss_omega_0}. One curve corresponds to the oscillation of the atoms of the first type and the second curve corresponds to the second type of atoms.  Time moment is $t=0.5$. \label{pic_out_phase_zoom}}
\end{figure}

On the figure \ref{pic_out_phase_sum} we present the sum of two waves --- acoustical and optical, where optical wave is presented by its envelope curves. We also presented on this graph the numerical solution. On the figure \ref{pic_out_phase_zoom} we presented the same waves separately in more fine scale.

\begin{figure}[!h]
\begin{center}
\includegraphics[scale=0.6]{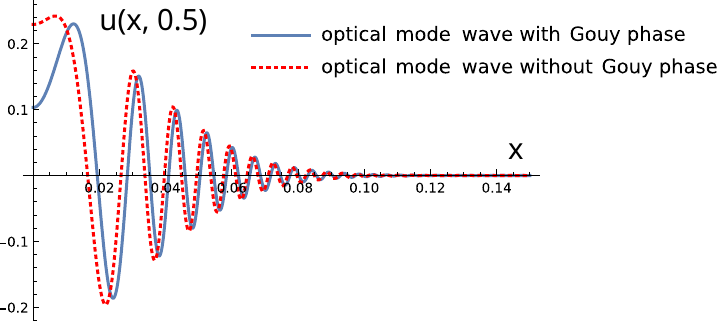}
\includegraphics[scale=0.6]{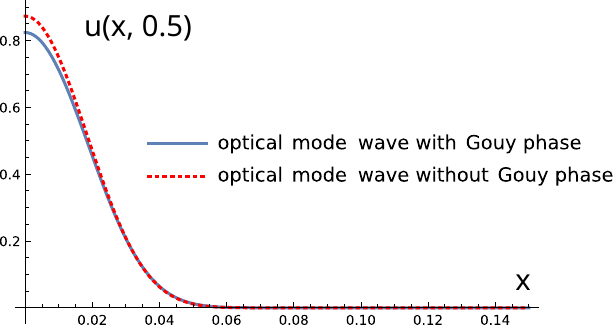}
\end{center}
\caption{On the right: the comparison of the function \eqref{opt_Gauss_omega_0} with and without the Gouy phase for the dispersive case. On the right: the function \eqref{opt_Gauss_omega_0} with and without Gouy phase for the non-dispersive case. Time moment is $t=0.5$. \label{pic_Gouy_phase}}
\end{figure}

Finally we want to demonstrate that the Gouy phase plays the significant role for the dispersive model of the optical mode. On the figure \ref{pic_Gouy_phase} we demonstrated function \eqref{opt_Gauss_omega_0} for two cases --- dispersive case and  the case of the long-wave approximation ($\mu=\sqrt{h}\approx 0.01679$.) We separately calculated the optical wave profile \eqref{opt_Gauss_omega_0}, setting the Gouy phase to zero and compare it with the asymptotic full formula.  We can see, that even in the long-wave approximation the Gouy phase plays some role, while in the dispersive case the Gouy phase starts to play important role.

Now let us consider the case of the sufficiently short perturbation. We choose the initial perturbation in the form of the Gaussian exponential $A(\xi)=e^{-\xi^2/2}$ and the zero initial velocity. Solving the equation \eqref{low_bound_mu_eq}, we arrive that $\mu\approx 0.0009$ which is $\mu\approx 3h$ and  $\delta\approx 1/3$.  

\begin{figure}[!h]
\begin{center}
\includegraphics[scale=0.7]{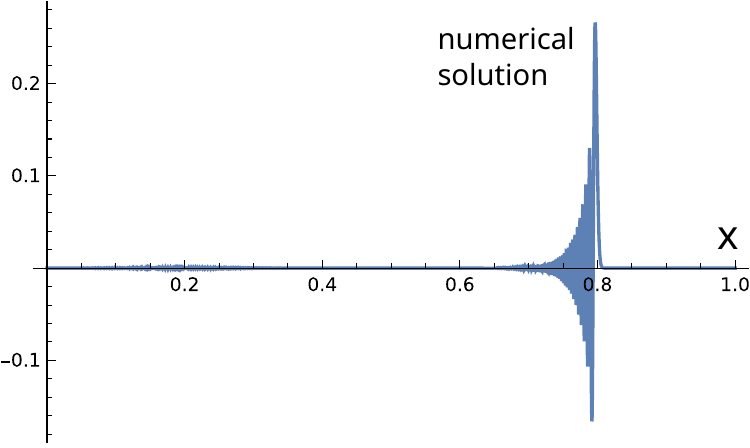}
\end{center}
\caption{Numerical solution of the system of equation \eqref{osc_lattice_dimens} for the case of the comparable width of perturbation and lattice step, $\mu\approx 3h=0.0009$. This part demonstrates the wave propagating to the right. \label{pic_storng_disp_num_sol_acoust}}
\end{figure}

 On the figure \ref{pic_storng_disp_num_sol_acoust} we presented the part of the numerical solution of the initial system of equations \eqref{osc_lattice_dimens} for this case. It represents the wave which propagates to the right. From this graph we can see, that for the presented situation the amplitude of the optical mode wave is negligible compare with the acoustical mode wave.

\begin{figure}[!h]
\begin{center}
\includegraphics[scale=0.58]{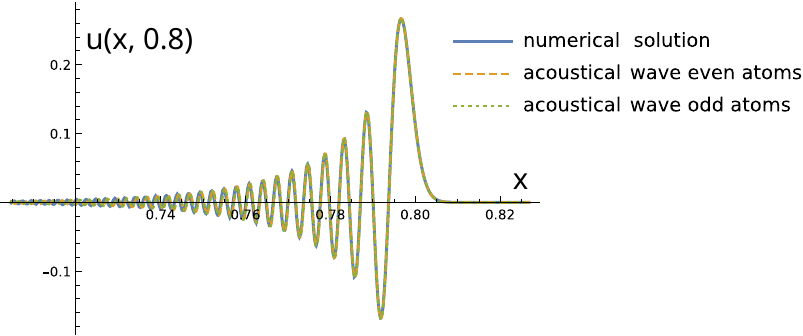}
\includegraphics[scale=0.58]{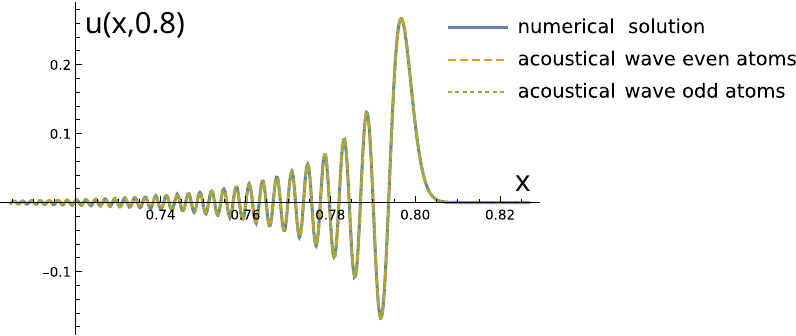}
\end{center}
\caption{On the left: graphs of the components of the vector-function \eqref{I12_Airy_strong} compared with the numerical solution. On the right: graphs of the components of the vector corresponding to the uniform asymptotic (\ref{I1_unif_right_wave_strong_disp}). Time moment is $t=0.8$. The solid line describes the numerical solution, the dashed and dotted lines describe the oscillations of the atoms of the even and odd types.  \label{pic_storng_disp_acoust}}
\end{figure}

 On the figure \ref{pic_storng_disp_acoust} we presented graphs of the acoustical wave branch, corresponding to the asymptotic representation near the wave front and uniform asymptotic representation, corresponding to the initial condition in the form of the Gaussian exponential. Visually one can tell, that the difference between the asymptotic near the front and the uniform asymptotic is in the tail of the wave when the principal waves have passed and only small oscillations are left.

\section{Conclusion}

Wave propagation in diatomic lattices stands as a fundamental problem in physics and mathematics, with far-reaching applications across composite materials, electrical transmission, and molecular dynamics. Yet despite its importance, the rigorous analysis of localized disturbances evolving in such systems has remained incomplete. This work addresses that gap through a comprehensive mathematical framework that bridges discrete lattice models and continuous descriptions while preserving the full physics at all length scales.

Our pseudo-differential operator approach represents a substantial advance over traditional long-wave approximations. By preserving the complete dispersion relations for both acoustic and optical modes rather than truncating to leading-order terms, we capture physically essential phenomena that were previously inaccessible to analytical treatment. This contrasts sharply with earlier methods that either yield ill-posed equations or introduce artificial mathematical complexity.

The formulation of the Cauchy problem emerges as a critical achievement of this work. By systematically determining initial conditions through Kotel'nikov-Whittaker-Shannon interpolation grounded in pseudo-differential theory, we establish a rigorous bridge between perturbations on the discrete lattice and continuous initial data. This methodology, surprisingly absent from the prior literature despite the model's fundamental status, provides a principled foundation for all subsequent analysis. 

When perturbation width significantly exceeds the lattice step, wave behavior reduces to the classical wave equation, validating long-wave  predictions. Reducing the perturbation width we arrive to the dispersive models, when perturbation width and lattice spacing are connected via the power law $3/2$. 

When the width of the perturbation is comparable with the lattice step, the dynamics is different. Dispersion effects become dominant, qualitatively altering wave propagation patterns and requiring the full pseudo-differential treatment. 

We derived explicit analytical asymptotic formulas for wave propagation across distinct physical regimes, expressed in terms of Airy functions and their derivatives. These formulas reveal previously unnoticed effects in diatomic crystal dynamics that emerge from the interplay between perturbation size and lattice spacing. The analytical formulas allow us to define the  Gouy time phase for the optical mode wave and construct the correct simple approximations for the pseudo-differential equations compare for each considered case.


Beyond theoretical rigor, our approach offers substantial practical benefits. Direct numerical integration of the original discrete system is computationally intensive, requiring calculations with large matrices. In contrast, our analytical formulas provide explicitly evaluated asymptotic expressions in terms of the Airy functions. The resulting computational speedup is dramatic: wave-front approximations can be evaluated instantaneously, enabling efficient parameter studies and design optimization in applications where repeated evaluations are necessary.

The methods developed here particularly the pseudo-differential operator framework and localized initial conditions parametrization - extend beyond diatomic systems. They provide a template for analyzing polyatomic lattices and more complex discrete structures where complete dispersion preservation is essential.

In conclusion, this work provides both conceptual clarity and practical tools for understanding wave dynamics in periodic media. By rigorously connecting microscopic discrete physics to macroscopic continuous descriptions while respecting the complete frequency spectrum, we establish a foundation for more sophisticated analyses of complex lattice systems. The resulting analytical formulas serve as efficient computational engines and analytical benchmarks against which numerical simulations can be validated - resources that will prove invaluable as research in, for example, phonon engineering and metamaterial design.

\appendix

\section{Proof of the Theorem \ref{thm_acoustic_I12_strong_disp}  }
\label{app_stron_disp}

Here we demonstrate how integral (\ref{I1_strong}), corresponding to the acoustical mode, can be asymptotically calculated with the help of the Airy function.

We have the following equality for the complex conjugation of the semi-discrete Fourier transform: $\widetilde{U}^0(-p)=\overline{\widetilde{U}^0(p)}$. The matrix $\Gamma^{1/2}\mathcal{P}_1(\delta p)\Gamma^{-1/2}$ is even and real-valued, as well as the function $\omega_1(\delta p)$. The function $\omega_1(\delta p)$ itself is non-smooth near $p=0$, therefore we are presenting the smooth and odd continuation of this function 
$$
\Omega(\delta  p)=
\begin{cases}
\omega_1(\delta p),\quad p\ge 0,\\
-\omega_1(- \delta p),\quad p\le 0.
\end{cases}
$$
With the help of this function we can define the following integrals 
\begin{gather}
\label{I11_strong_disp}
I^{ac}_{1}(x,\,t)=\frac{\delta}{2\pi}\int\limits_{-\pi/(2\delta)}^{\pi/(2\delta)}\Gamma^{1/2}\mathcal{P}_1(\delta p)\Gamma^{-1/2}\widetilde{U}^0(p)e^{\frac{i}{\mu}(px+\frac{1}{\delta}\Omega(\delta p)t)}dp,\\
\label{I12_strong_disp}
I^{ac}_{2}(x,\,t)=\frac{\delta}{2\pi}\int\limits_{-\pi/(2\delta)}^{\pi/(2\delta)}\Gamma^{1/2}\mathcal{P}_1(\delta p)\Gamma^{-1/2}\widetilde{U}^0(p)e^{\frac{i}{\mu}(px-\frac{1}{\delta}\Omega(\delta p)t)}dp.
\end{gather}
We have the following equality
$$
Re\, I^{ac}(x,\,t)=\frac{1}{2}(I^{ac}(x,\,t)+\overline{I^{ac}(x,\,t)})=Re\, I^{ac}_{1}(x,\,t)+Re\, I^{ac}_{2}(x,\,t).
$$

The integrals (\ref{I11_strong_disp}), \eqref{I12_strong_disp} describe two waves, where $I^{ac}_{1}(x,\,t)$ is propagating to the left, and $I^{ac}_{2}(x,\,t)$ is propagating to the right.
They can be presented as a WKB exponential outside of the wave fronts $x=\pm c_0t$ with the help of the stationary phase method. The wave fronts correspond to the singular stationary point $p=0$, where the second derivative is zero $\omega_1''(0)=0$. Near the wave front the integrals can be evaluated with the help of the Airy function and its derivative.

Here we present detailed calculations for the integral \eqref{I12_strong_disp}, for the integral \eqref{I11_strong_disp} the calculations are similar and we will introduce them briefly.

We introduce the smooth even cutoff function $\chi(p)$ near the point $p=0$ and rewrite the integral $I^{ac}_{2}(x,\,t)$ as the sum
\begin{gather}
\label{big_disp_int_ac_separ_zero}
I^{ac}_{2}(x,\,t)=\frac{\delta}{2\pi}\int\limits_{-\pi/(2\delta)}^{\pi/(2\delta)}\chi(p)\Gamma^{1/2}\mathcal{P}_1(\delta p)\Gamma^{-1/2}\widetilde{U}^0(p)e^{\frac{i}{\mu}(px-\frac{1}{\delta}\Omega(\delta p)t)}dp+\\
\nonumber
+\frac{\delta}{2\pi}\int\limits_{-\pi/(2\delta)}^{\pi/(2\delta)}(1-\chi(p))\Gamma^{1/2}\mathcal{P}_1(\delta p)\Gamma^{-1/2}\widetilde{U}^0(p)e^{\frac{i}{\mu}(px-\frac{1}{\delta}\Omega(\delta p)t)}dp.
\end{gather}

\paragraph{Proof of the Statement \ref{st_I11_strong_disp}.} The wavefront $x=c_0 t$  corresponds to the point $p=0$ and for the asymptotic near the wave front we need integration over vicinity of the $p=0$.

Near the point $p=0$, we have the  expansion $\Omega(\delta p)=c_0\delta p-q\delta^3p^3/3+O(p^5)$. If the cutoff function $\chi(p)$ covers the $O(\mu^{1/3})$ area of the point $p=0$, then we can substitute this expansion into the integral $I^{ac}_{2}(x,\,t)$.
We can also expand the area of integration over all real line, which gives the integral
\begin{equation}
\label{I0_wave_front_srong_disp}
I^{ac}_{2}(x,\,t)=\frac{\delta}{2\pi}\int\limits_{-\infty}^{\infty}\chi(p)\Gamma^{1/2}\mathcal{P}_1(\delta p)\Gamma^{-1/2}\widetilde{U}^0(p)e^{\frac{i}{\mu}((x-c_0t)p+qt \delta^2\frac{p^3}{3})}dp(1+O(\mu^{2/3}))
\end{equation}

The main idea to calculate the integral (\ref{I0_wave_front_srong_disp}) is to use of the H\"{o}ramnder's formula, similar to  \cite{dobrokhotov_maslovs_2014}. Let us demonstrate this approach for the Airy function. Airy function satisfies the equation $v''(z)-z v(z)=0$, and, by its definition (\ref{Airy_func_def}), for $z\in\mathbb{R}$ we have
$$
\frac{1}{2\pi}\int\limits_{\mathbb{R}}(-t^2) e^{i (z t+\frac{t^3}{3})}dt=\frac{d^2}{dz^2}\frac{1}{2\pi}\int\limits_{\mathbb{R}} e^{i (z t+\frac{t^3}{3})}dt=
z\frac{1}{2\pi}\int\limits_{\mathbb{R}} e^{i (z t+\frac{t^3}{3})}dt=z Ai(z).
$$
This means that we can replace the multiplication by $(-t^2)$ of the integrand by the multiplication on the argument of the Airy function.

The integral (\ref{I0_wave_front_srong_disp}) is of the similar form and the H\"{o}rmander's formula allows us to replace some function $f(p^2)$  near the point $p=0$ by other function $f(x,\,t)$  and move the corresponding argument outside of the integral.

For the phase function in \eqref{I0_wave_front_srong_disp} we have
$$
x-c_0t+qt\delta^2 p^2=0\Rightarrow p^2=-\frac{x-c_0t}{q\delta^2t}.
$$

We have the following equality $\widetilde{U}^0(p)=\widetilde{U}^0_{1}(p^2)+p\widetilde{U}^0_{2}(p^2)$, where functions $\widetilde{U}^0_{1,2}$ are defined in (\ref{I12_cor_V1_V2_strong_disp}).
After that we can replace $p^2$ by the fraction above and present the integral (\ref{I0_wave_front_srong_disp}) as the following combination
\begin{gather}
\label{I12_comb_int_strong_disp}
I^{ac}_{2}(x,\,t)=
\Gamma^{1/2}\mathcal{P}_1(0)\Gamma^{-1/2}  Re\,\Bigl(\widetilde{U}^0_{1}\left(-\frac{x-c_0t}{q\delta^2t}\right)\frac{\delta}{2\pi}\int\limits_{-\infty}^{\infty}e^{\frac{i}{\mu}((x-c_0t)p+qt \delta^2\frac{p^3}{3})}dp+\\
\nonumber
+\widetilde{U}^0_{2}\left(-\frac{x-c_0t}{q\delta^2 t}\right)\frac{\delta}{2\pi}\int\limits_{-\infty}^{\infty}pe^{\frac{i}{\mu}((x-c_0t)p+q\delta^2 t \frac{p^3}{3})}dp\Bigr)(1+O(\mu^{2/3})).
\end{gather}
Because $\Gamma^{1/2}\mathcal{P}_1(\delta p)\Gamma^{-1/2}$ is even, we used the expansion $\Gamma^{1/2}\mathcal{P}_1(\delta p)\Gamma^{-1/2}=\Gamma^{1/2}\mathcal{P}_1(0)\Gamma^{-1/2}+O(p^2)$ which gives the correction $O(\mu^{2/3})$. 

The first integral in (\ref{I12_comb_int_strong_disp}) can be presented with the help of the Airy function and the second one is the derivative. After the change of variables $\eta=(qt/\mu)^{1/3}p$ we have 
\begin{equation}
\label{Ai_int}
\frac{\delta}{2\pi}\int\limits_{-\infty}^{\infty}e^{\frac{i}{\mu}((x-c_0t)p+q\delta^2 t \frac{p^3}{3})}dp=
\left(\frac{\mu\delta}{qt}\right)^{1/3} Ai(y(x,\,t)),\quad y(x,\,t)=\frac{x-c_0t}{\mu^{2/3}(q\delta^2 t)^{1/3}}.
\end{equation}
For the second integral in (\ref{I12_comb_int_strong_disp}) we have the following equality
\begin{equation}
\label{Ai_Prime_int}
\frac{\delta}{2\pi}\int\limits_{-\infty}^{\infty}pe^{\frac{i}{\mu}((x-c_0t)p+q\delta^2 t \frac{p^3}{3})}dp=
-i\mu\frac{\partial}{\partial x}\frac{\delta}{2\pi}\int\limits_{-\infty}^{\infty}e^{\frac{i}{\mu}((x-c_0t)p+q\delta^2 t \frac{p^3}{3})}dp=- i\frac{1}{\delta^{1/3}}\left(\frac{\mu}{qt}\right)^{2/3}Ai'(y(x,\,t)).
\end{equation}

Since $p$ belongs to the $O(\mu^{1/3})$-area of the point $p=0$, we have $|x-c_0t|=O(p^2)=O(\mu^{2/3})$. 
Thus in the $O(\mu^{2/3})$ vicinity of the wave front $x=c_0t$ we have the representation (\ref{I12_Airy_strong}). 

The similar technique is applied to the integral \eqref{I11_strong_disp}. Separation of the integration near point $p=0$ leads to the Airy representation of the integral in the in the $O(\mu^{2/3})$-vicinity of the wave front $x=-c_0t$
This finalizes the proof the Statement \ref{st_I11_strong_disp}.

\paragraph{Proof of the theorem \ref{thm_acoustic_I12_strong_disp}.} Let us start from the second integral in \eqref{big_disp_int_ac_separ_zero}, which corresponds to the integration outside of the point $p=0$. This integral can be calculated with the help of the stationary phase method.
Let us denote the phase of the integral 
$$
\Phi(p;\,x,\,t)=xp-\frac{1}{\delta}\Omega(\delta p)t.
$$
The stationary points  of this phase are defined by the equation
$$
\frac{\partial}{\partial p}\Phi(p;\,x,\,t)=x-\Omega'(\delta p) t=0.
$$
When $0\le x<c_0 t$, this equation has two solutions $\pm p_2=\pm p_2(x,\,t)$ where $p_2(x,\,t)>0$. Moreover, this equation has solutions only when $0\le x\le c_0t$, because $0\le \Omega'(\delta p)\le c_0$. For any other values of $x$ this integral is of order $O(\mu^\infty)$.

In stationary points $\pm p_2$ the second derivative is non-zero
$$
\frac{\partial^2}{\partial p^2}\Phi(\pm p_2;\,x,\,t)=-\delta\Omega''(\pm \delta p_2)t\not=0\Leftrightarrow \Omega''(\pm \delta p_2)\not=0.	
$$
We have inequalities $\Omega''(\delta p)\le 0$ when $p\in[0,\,\pi/(2\delta)]$ and $\Omega''(\delta p)\ge 0$ when $p\in[-\pi/(2\delta),\,0]$. 

So, in the case when $0\le x<c_0t$, by the stationary phase method we have the WKB representation for the integral $I^{ac}_{2}(x,\,t)$ outside of the wave front $x=c_0t$. Passing back from the function $\Omega(\delta p)$ to the function $\omega_1(\delta p)$ we get the following
\begin{align}
\label{I12_WKB_stong_disp}
I^{ac}_{2}(x,\,t)=\sqrt{\frac{\delta  \mu}{2\pi}}\frac{\Gamma^{1/2}\mathcal{P}_1(\delta p_2)\Gamma^{-1/2}}{\sqrt{|\omega_1''(\delta p_2)|}\sqrt{t}}\Biggl(\widetilde{U}^0(p_2)e^{\frac{i}{\mu}(p_2 x-\frac{1}{\delta}\omega_1(\delta  p_2)t)+i\frac{\pi}{4}}
+\widetilde{U}^0(-p_2)e^{-\frac{i}{\mu}(p_2 x-\frac{1}{\delta}\omega_1(\delta p_2)t)-i\frac{\pi}{4}}\Biggr)(1+O(\mu)).
\end{align}

In order to construct the uniform asymptotic representation on the interval $x\in[0,\,c_0t+\varepsilon)$ for some $\varepsilon>0$, we need to introduce the following two functions \cite{anikin_uniform_2019}
$$
A^\pm(y)=\sqrt{\pi}\left[\left(\frac{3}{2}y\right)^{1/6}Ai\left(-\left(\frac{3}{2}y\right)^{2/3}\right)\pm i\left(\frac{3}{2}y\right)^{-1/6}Ai'\left(-\left(\frac{3}{2}y\right)^{2/3}\right)\right].
$$
These functions have the asymptotic expansion for large $y$
\begin{equation}
\label{func_A_WKB_as_strong_disp}
A^\pm(y)=e^{\pm i(y-\pi/4)}\left(1+O\left(\frac{1}{y}\right)\right),\quad y\to +\infty.
\end{equation}
Using this asymptotic expansion, we replace the exponents in the WKB representations (\ref{I12_WKB_stong_disp}) 
by the corresponding functions $A^\pm(y)$ with the proper argument. 

Let us define the function
$$
S_2(p_2)=-(p_2 x-\frac{1}{\delta}\omega_1(\delta p_2)t).
$$
Then for $x\in [0,\,c_0t+\varepsilon)$, $\varepsilon>0$, we have the representation
\begin{equation}
\label{I12_A_func_strong_disp}
I^{ac}_{2}(x,\,t)=\sqrt{\frac{\mu \delta}{2\pi}}\frac{\Gamma^{1/2}\mathcal{P}_1(\delta p_2)\Gamma^{-1/2}}{\sqrt{|\omega_1''(\delta p_2)|}\sqrt{t}}\left(\widetilde{U}^0(p_2)A^-\left(\frac{S_2(p_2)}{\mu}\right)+\widetilde{U}^0(-p_2)A^+\left(\frac{S_2(p_2)}{\mu}\right)\right)(1+O(\mu^{2/3})).
\end{equation}
 Function $p_2=p_2(x,\,t)\ge 0$ is the positive solution of the equation $ x-\omega_1'(\delta p_2)t=0$ when $0\le x<c_0t$. If  $x\in[0,\,c_0t)$ and is outside of the vicinity of the leading edge $x=ct$, then by (\ref{func_A_WKB_as_strong_disp}) we immediately obtain  the WKB representiation (\ref{I12_WKB_stong_disp}) from (\ref{I12_A_func_strong_disp}).

After combining of the common multipliers of Airy function and its derivative in (\ref{I12_A_func_strong_disp}) we see, that it is exactly the desired representation (\ref{I1_unif_right_wave_strong_disp}).

Let us now explain in details why the formulae (\ref{I1_unif_right_wave_strong_disp}) and (\ref{I12_Airy_strong}) coincide,  in principle term, near the wave front $x=c_0t$.

Near the point $x=c_0t$ we have the approximation 
$p_2(x,\,t)=\sqrt{(c_0t -x)/q\delta^2t}$.
The Taylor expansion near $p=0$ gives
$$
S_2(p_2)=(x-c_0t)p_2-q\delta^2t\frac{p_2^3}{3}+O(p_2^5)\Rightarrow
S_2(p_2(x,\,t))=\frac{2}{3}\frac{(c_0t-x)^{3/2}}{\sqrt{q\delta^2 t}}+O\left(\left(\frac{c_0t -x}{q\delta^2t}\right)^{5/2}\right).
$$
From this we can conclude that, when $x\to c_0t$, we have the limit
$$
-\left(\frac{3}{2}\frac{S(p_2)}{\mu}\right)^{2/3}\to \frac{x-c_0t}{\mu^{2/3}(q\delta^2 t)^{1/3}}.
$$
We see that the argument of the Airy function converges to the same argument of the Airy function in the representation (\ref{I12_Airy_strong}).

Now let us analyze the amplitude of the Airy function. We have the following approximations when $x$ near the wave front 
$$
\frac{1}{\delta}\omega_1''(\delta p_2) t=-2p_2q\delta^2 t= -2q\delta^2 t \sqrt{\frac{c_0t -x}{q\delta^2 t}}=-2 \sqrt{q\delta^2 t}\sqrt{c_0t-x},\quad 
\left(\frac{3}{2}\frac{S_2(p_2)}{\mu}\right)^{1/6}= \frac{1}{\mu^{1/6}}\frac{(c_0t-x)^{1/4}}{(q\delta^2 t)^{1/12}}.
$$
Therefore for the amplitude, when $x\to c_0t$, we have the limit
\begin{gather*}
\frac{\sqrt{2\mu \delta}}{\sqrt{|\omega_1''(\delta p_2)|}\sqrt{ t}}\left(\frac{3}{2}\frac{S_2(p_2)}{\mu}\right)^{1/6}\to\frac{\mu^{1/3}\delta^{1/3}}{(qt)^{1/3}},\quad
\frac{p_2\sqrt{2\mu}  }{\sqrt{|\omega_1''(\delta p_2)|}\sqrt{t}} \left(\frac{3}{2}\frac{S(p_2)}{\mu}\right)^{-1/6}\to  \frac{\mu^{2/3}\delta^{2/3}}{(q t)^{2/3}}.
\end{gather*}
This shows that the asymptotic formulas \eqref{I12_Airy_strong} and \eqref{I1_unif_right_wave_strong_disp} in principal terms coincide.

Similarly we have the representation of the integral (\ref{I11_strong_disp}).
When $x<-c_0t$ or $x>0$, the integral $I^{ac}_{1}(x,\,t)=O(\mu^\infty)$. The phase of this integral is $\Phi(p;\,x,\,t)=xt+\Omega(\delta p)t/\delta$ and the equation for the stationary point is $\Phi'(p;\,x,\,t)=x+\Omega'(\delta p)t=0$.
This equation has two solutions $\pm p_1=\pm p_1(x,\,t)$ when $-c_0t<x\le 0$. The signature of the second derivative $\Phi''(\pm p_1;\,x,\,t)$ coincides with the sign of the $\Omega''(\pm \delta p_1)$. This gives us the WKB representation of the integral $I^{ac}_{1}(x,\,t)$ outside of the singular point $p=0$.

Let us define the function
$$
S_1(p_1)=p_1 x+\frac{1}{\delta}\omega_1(\delta p_1)t.
$$
Then for some $\varepsilon>0$ when $x\in(-c_0t-\varepsilon,\,0]$ the following representation holds
$$
I^{ac}_{1}(x,\,t)=\sqrt{\frac{\mu}{2\pi}}\frac{\Gamma^{1/2}\mathcal{P}_1(\delta p_1)\Gamma^{-1/2}}{\sqrt{|\omega_1''(\delta p_1)|}\sqrt{t}}\left(\widetilde{U}^0(p_1)A^+\left(\frac{S_1(p_1)}{\mu}\right)+\widetilde{U}^0(-p_1)A^-\left(\frac{S_1(p_1)}{\mu}\right)\right)(1+O(\mu^{2/3})).
$$
Here function $p_1=p_1(x,\,t)\ge 0$ is the positive solution of the equation $x+\omega_1'(p_1)t=0$ when $-c_0t<x\le 0$. Similar reasoning leads for the representation \eqref{I11_Airy_strong_disp}.

This finalizes the proof of the Theorem \ref{thm_acoustic_I12_strong_disp}.


\bibliographystyle{plainurl}      
\bibliography{crystal_24_07_2026}

\end{document}